\documentclass[journal]{IEEEtran}
\usepackage[utf8]{inputenc}
\usepackage[english]{babel}
\usepackage{hyperref}
\usepackage{graphicx}
\usepackage{amsmath}
\usepackage{subfig}
\usepackage{caption}
\usepackage{textcomp}
\usepackage{stfloats}
\usepackage{url}
\usepackage{verbatim}
\usepackage{graphicx}
\usepackage{cite}
\graphicspath{{./Imagenes/}}
\usepackage{multicol}
\usepackage{amsmath, lipsum}
\usepackage{cuted}
\usepackage{enumitem}
\begin{document}

\title{Improved PCRLB for radar tracking in clutter with geometry-dependent target measurement uncertainty and application to radar trajectory control}

\author{Yifang~Shi,~\IEEEmembership{Member, IEEE,} Yu Zhang, Linjiao Fu, Dongliang Peng,  Qiang Lu, \IEEEmembership{Member, IEEE,} Jee Woong Choi and Alfonso Farina,~\IEEEmembership{Life Fellow, IEEE}

\textit{This work has been submitted to the IEEE for possible publication. Copyright may be transferred without notice, after which this version may no longer be accessible.}

\thanks{Yifang Shi, Yu Zhang, Linjiao Fu, Qiang Lu and Dongliang Peng are with the School of Automation, Hangzhou Dianzi University, Hangzhou, 310018, China (e-mail: syf2008@hdu.edu.cn, a1137725141@163.com, fulj121@163.com, dlpeng@hdu.edu.cn, Lvqiang@hdu.edu.cn)}
\thanks{Jee Woong Choi is with the department of Marine Science and Convergence Engineering, Hanyang University, Ansan, 15588, Republic of Korea (e-mail:choijw@hanyang.ac.kr)}
\thanks{Alfonso Farina is with Selex-es (retired), consultant,, Roma, 00144, Italy (e-mail:alfonso.farina@outlook.it)}
\thanks{This work was supported in part by the
National Natural Science Foundation of China under Grant 62471165 and 61901151; in part by the Zhejiang Provincial Natural Science Foundation of China under Grant LZ23F030002. (Corresponding author: Dongliang Peng.)}}


\maketitle
\begin{abstract}
In realistic radar tracking, target measurement uncertainty (TMU) in terms of both detection probability and measurement error covariance is significantly affected by the target-to-radar (T2R) geometry. However, existing posterior Cramér-Rao Lower Bounds (PCRLBs) rarely investigate the fundamental impact of T2R geometry on target measurement uncertainty and eventually on mean square error (MSE) of state estimate, inevitably resulting in over-conservative lower bound. To address this issue, this paper firstly derives the generalized model of target measurement error covariance for bistatic radar with moving receiver and transmitter illuminating any type of signal, along with its approximated solution to specify the impact of T2R geometry on error covariance. Based upon formulated TMU model, an improved PCRLB (IPCRLB) fully accounting for both measurement origin uncertainty and geometry-dependent TMU is then re-derived, both detection probability and measurement error covariance are treated as state-dependent parameters when differentiating log-likelihood with respect to target state.  Compared to existing PCRLBs that partially or completely ignore the dependence of target measurement uncertainty on T2R geometry, proposed IPCRLB provides a much accurate (less-conservative) lower bound for radar tracking in clutter with geometry-dependent TMU. The new bound is then applied to radar trajectory control to effectively optimize T2R geometry and exhibits least uncertainty of acquired target measurement and more accurate state estimate for bistatic radar tracking in clutter, compared to state-of-the-art trajectory control methods.  

\end{abstract}


\section{Introduction}

\IEEEPARstart{T}{he} Cramér-Rao Lower Bound (CRLB) is defined as the inverse of Fisher information matrix (FIM) that quantitatively evaluates the maximum information extracted from existing knowledge about the unknown vector \cite{ref:1}. In the context of radar tracking, since the target kinematic state to be estimated is a random vector and usually time-varying, the posterior CRLB (PCRLB) is then defined and calculated in a Riccati-like recursion manner \cite{ref:2}, providing the lowest mean-square error (MSE) bound that any unbiased estimators can achieve in theory. In general, the performance of radar tracking in clutter is dominated by the radar measurement uncertainty that lies on twofold: the measurement origin uncertainty (MOU) due to the presence of clutter and multi-target, and the target measurement uncertainty stemming from random miss-detection and statistical noise corruption. Open literatures are dedicated to extend the preliminary PCRLB in \cite{ref:2} for radar tracking with different levels of radar measurement uncertainties. Such as \cite{ref:6,ref:7,ref:8,ref:9},\cite{ref:15} introduced the information reduction factor (IRF) into PCRLB to account for the MOU caused by clutter disturbance,  \cite{ref:10,ref:11,ref:12} derived the PCRLB for multi-radar tracking problem by further considering the MOU caused by the presence of multitarget, \cite{ref:16} derived the PCRLB for distributed range-only localization with MOU and signal-noise-ratio (SNR)-dependent target detection probability, more recently, \cite{ref:17} derived the exact
FIM (EFIM) with the MOU and state-dependent target detection
probability, just name a few. However, it is noteworthy that existing literature formulates PCRLB for radar tracking with radar measurement uncertainty under a basic assumption: the target measurement error covariance is independent of the target-to-radar (T2R) geometry. This assumption obviously deviates from the realistic radar operation \cite{ref:17, ref:19} and inevitably results in over-conservative PCRLBs. Particularly, in bistatic radar system, the TMU in terms of both target detection probability and measurement error covariance are apparently affected by the T2R triangular geometry, which eventually determines the tracking performance \cite{ref:20}. 

Motivated by the fact that the target measurement uncertainty (TMU) is highly affected by the T2R geometry, along with the fact that radar platform is usually collaborative and capable of maneuvering (e.g., airborne), existing literatures pay tremendous attention to optimize the T2R geometry via controlling the trajectory of radar platform. In such, the radar enables to cognitively adapt its moving path to acquire higher quality target measurement based on perceived environment and eventually improve the tracking performance. The radar trajectory control is essentially a nonlinear control problem that aims to assign radar best state at each sampling time under constraints on platform maneuver limitations (such as maximum acceleration and angular rate). To achieve radar trajectory control in real-time, existing literatures concentrate on the partially observed Markov Decision Process (POMDP) framework. This framework formulates the control decision-making problem as a first-order Markov process over time, and selects the best control commands via optimizing an objective function that quantitatively measures the goodness on the trajectory-control resulted estimation. This objective function can be either formulated into an information-driven reward as in \cite{ref:21,ref:22,ref:23,ref:24}, or a task-driven cost as in \cite{ref:26,ref:27,ref:29,ref:30,ref:31,ref:32}. The optimal control commands are then obtained by either maximizing reward function or minimizing cost function. Nevertheless, it is noteworthy that both the reward and cost function in existing literatures necessities to implement pseudo tracking using predicted ideal measurement set (PIMS) that is generated by traversing all admissible radar trajectory control commands, and calculate posterior state estimates before actuating radar to its eventual state. This exhaustive enumeration operation inevitably leads to cumbersome computation. Furthermore, existing radar trajectory control literatures rarely investigate the fundamental impact of T2R geometry on the TMU and tracking accuracy, which in turn suppresses the benefits of radar trajectory optimization.

 In this paper, we pay specific interests on radar tracking in clutter with T2R geometry-dependent TMU, and rigorously formulate the impact of T2R geometry on TMU and MSE lower bound. Three main contributions of this paper are summarized as follows:
\begin{itemize}
    \item Derive the generalized form of target measurement error covariance for bistatic radar with moving receiver and transmitter illuminating any type of signal, and also give its approximated solution to specify the impact of T2R geometry on error covariance. 
    \item Rigorously derive an improved  PCRLB (IPCRLB) for radar tracking in clutter by fully accounting for radar measurement uncertainty in terms of both MOU and geometry-dependent TMU. The information gain factor (IGF) is specifically defined to quantify additional target Fisher information arising from considering dependence of target detection probability and measurement error covariance on T2R geometry. This factor reduces to zero when the dependence is completely ignored. The derived IPCRLB provides a much more accurate (less conservative) MSE lower bound compared to state-of-the-art bounds that partially or completely ignore the dependence of TMU on T2R geometry. Numerical investigation demonstrates the improvement of IPCRLB over existing bounds tends to enlarge as TMU increases.
    \item  To leverage the superior IPCRLB to radar trajectory control, the trace of predictive IPCRLB is proposed as the cost function and minimized to optimize T2R geometry to acquire high-quality target measurement for improved radar tracking in clutter. Since IPCRLB fundamentally investigates the impact of T2R
geometry on the MSE lower bound, compared to state-of-the-art control methods, proposed method optimizes T2R geometry more effectively and exhibits least uncertainty of acquired target measurement and more accurate tracking results.
\end{itemize}

The organization of this paper: geometry-dependent TMU is formulated in section~\ref{sec:2}. section~\ref{sec:3} presents the detailed derivation of the proposed IPCRLB fully accounting for radar measurement uncertainty, followed by its application to radar trajectory control in section~\ref{sec:4}, section~\ref{sec:5} validates proposed model and methods with simulation study, closed with conclusion in section~\ref{sec:5}.

\section{Geometry-dependent target measurement uncertainty} \label{sec:2}
The bistatic radar system consists of separately deployed transmitter and receiver, in which transmitter is dedicated to transmit certain type of signal towards to surveillance area, while the receiver is responsible for collecting and processing reflected signals. Either the transmitter or the receiver can be mounted to moving platform and is usually collaborative. The bistatic radar is a highly generalized radar system that coincides to be monostatic radar when the transmitter and receiver are collocated, or becomes a passive radar when the target serves as the illuminator. Without loss of generality, this section uses bistatic radar to demonstrate the impact of T2R geometry on the TMU. Unlike \cite{ref:18}, \cite{ref:33,ref:34,ref:35}, which are limited to static radar with specific signals, the geometry-dependent bistatic measurement error covariance formulated here is a much more generalized form. It can be used for bistatic radar tracking with moving receiver and transmitter illuminating any type of signal.
\subsection{Bistatic radar overview}
\begin{figure}[h!] 
\centering
\includegraphics[width=0.45\textwidth]{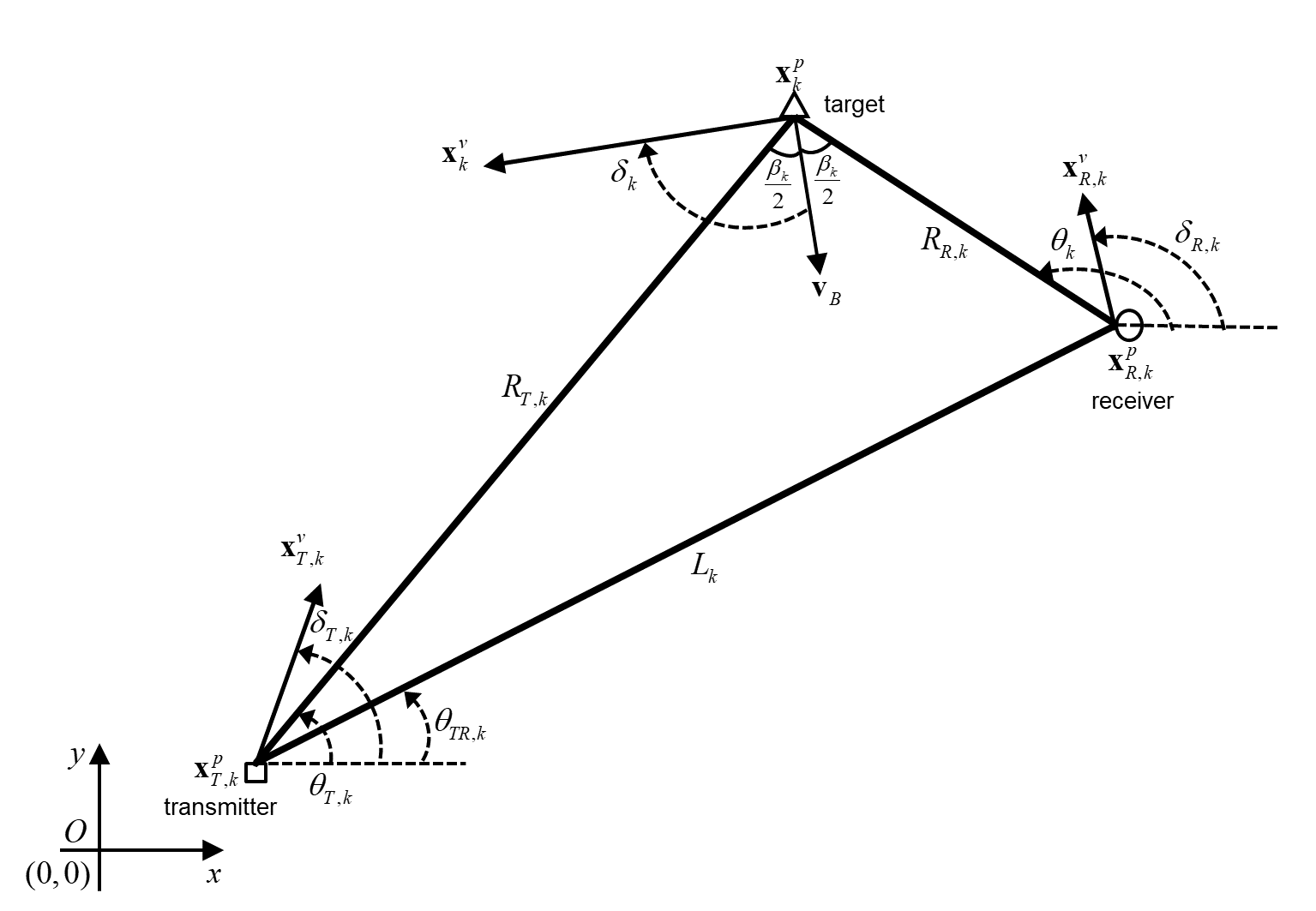}
\caption{Bistatic radar operation in 2D Cartesian coordinate}
\label{fig:BisGeometry}
\end{figure}

Instead of the specified north-referenced coordinate system \cite{ref:18}, \cite{ref:33}-\cite{ref:35}, we consider a more generalized two-dimensional (2D) Cartesian coordinate system to demonstrate bistatic radar operation for radar tracking with moving target, transmitter and receiver as shown in Fig.\ref{fig:BisGeometry}. The full state of either moving target, transmitter or receiver composites of two-dimensional position and velocity,  denoted by ${{\bf{x}}_k} = {\left[ {\left( {{\bf{x}}_k^p} \right)^T {\kern 5pt} {{\left( {{\bf{x}}_k^v} \right)}^T}} \right]^T}$, ${{\bf{x}}_{T,k}} = {\left[ {\left( {{\bf{x}}_{T,k}^p} \right)^T {\kern 5pt} {{\left( {{\bf{x}}_{T,k}^v} \right)}^T}} \right]^T}$ and ${{\bf{x}}_{R,k}} = {\left[ {\left( {{\bf{x}}_{R,k}^p} \right)^T {\kern 5pt} {{\left( {{\bf{x}}_{R,k}^v} \right)}^T}} \right]^T}$, respectively.  At each time $t_k$, the separately deployed target, transmitter and receiver form a bistatic triangular geometry that can be denoted by the abbreviation ${\Delta _k} \buildrel \Delta \over = \Delta \left( {{\bf{x}}_k^p,{\bf{x}}_{T,k}^p,{\bf{x}}_{R,k}^p} \right)$. As shown in Fig.\ref{fig:BisGeometry}, ${\Delta _k}$ can be described by a set of triangle parameters $\left({{L_k},{R_{T,k}},{R_{R,k}},{\theta _{RT,k}},{\theta _{T,k}},{\theta _k}} \right)$.  ${L_k}$ is the baseline range between transmitter and receiver, ${R_{T,k}}$ is the transmitter to target range, ${R_{R,k}}$ is the receiver to target range, with the target bistatic range defined by $d_k=R_{R,k}+R_{T,k}$. ${\theta _{TR,k}}$ is the angle of transmitter-to-receiver line-of-sight (LOS),  ${\theta _{T,k}}$ and ${\theta _k}$ are the transmitter and receiver look angles respectively, both measured positive counter-clockwise from the x-axis and restricted to $\left[ {0,2\pi } \right]$.  Among parameters above, ${\theta _{TR,k}}$ is used to locate the triangle in the 2D Cartesian space, while $\left( {{d_k},{L_k},{\theta _{T,k}},{\theta _k}}\right)$ is used to uniquely determine the shape of the triangle. Obviously, this triangle ${\Delta _k}$ is time-varying that changes as any of target, transmitter and receiver moves along its trajectory, and converges to be a straight line when the target-transmitter-receiver becomes collinear or transmitter-receiver is collocated. 

In bistatic radar system, the receiver cooperates with the transmitter to measure the time delay ${\tau _k}$ and Doppler shift ${\xi _k}$ of received target echo, which are then converted to obtain the bistatic range ${d_k}$ and bistatic velocity (bistatic range rate) ${v_k}$, while the direction-of-arrival (DOA) of target echo can be directly obtained as the receiver looking angle ${\theta _k}$. Therefore, the target state ${{\bf{x}}_k}$ can be  recursively estimated based on the bistatic measurement ${{\bf{z}}_k} = {\left[ {{d_k}{\kern 5pt} {v_k}{\kern 5pt} {\theta _k}} \right]^T}$. Relationship between bistatic measurement and target's state can be described by
\begin{equation}\label{eq:MeaEq}
{{\bf{z}}_k} = {\bf{h}}({{\bf{x}}_k}) + {{\bf{w}}_k},
\end{equation}
where ${\bf{h}}({{\bf{x}}_k})$ is a time-invariant nonlinear vector function of target state ${{\bf{x}}_k}$, and composites of noise-free bistatic range, bistatic velocity and DOA given by
\begin{subequations} \label{eq:MeaEq_Comp}
\small
\begin{align}
& {h^d}({{\bf{x}}_k}) = \left\| {{\bf{x}}_k^p - {\bf{x}}_{T,k}^p} \right\|{_2} + \left\| {{\bf{x}}_k^p - {\bf{x}}_{R,k}^p} \right\|{_2}\\
& {h^v}({{\bf{x}}_k}) = \frac{{{{({\bf{x}}_k^p - {\bf{x}}_{T,k}^p)}^T}\left( {{\bf{x}}_k^v - {\bf{x}}_{T,k}^v} \right)}}{{\left\| {{\bf{x}}_k^p - {\bf{x}}_{T,k}^p} \right\|{_2}}} + \frac{{{{({\bf{x}}_k^p - {\bf{x}}_{R,k}^p)}^T}\left( {{\bf{x}}_k^v - {\bf{x}}_{R,k}^v} \right)}}{{\left\| {{\bf{x}}_k^p - {\bf{x}}_{R,k}^p} \right\|{_2}}}\\
& {h^\theta }({{\bf{x}}_k}) = {\tan ^{ - 1}}\frac{{{y_k} - {y_{R,k}}}}{{{x_k} - {x_{R,k}}}},
\end{align}
\end{subequations}
where $\left\| . \right\|{_2}$ is the Euclidean norm, ${{\bf{w}}_k}$ denotes the additive white Gaussian noise. The relationship between bistatic signal $\left( {{\tau _k}, {\xi _k}} \right)$  and bistatic measurement $\left( {{d_k}, {v_k} } \right)$ can be described as follow 
\begin{equation} \label{eq:SigToMea}
{\tau _k} = \frac{{{d_k}}}{c},{\rm{ }}{\xi _k} = \frac{{{f_c}}}{c}{v_k}
\end{equation}
where $c$ and $f_c$ are the speed of light and carrier frequency of transmitted signal, respectively. Since the derivation of T2R geometry-dependent error covariance in section \ref{GeoDep-TMU} needs to calculate the derivative of ${\xi_k}$ with respect to ${d_k}$,  an alternative to formulate the Doppler shift ${\xi _k}$ as a function of bistatic range ${d_k}$ is given by (derivation refers to Appendix A in supplementary material)
\begin{equation} \label{eq:xi_dk}
\resizebox{.9\hsize}{!}{$
\begin{split}
& {\xi _k}  =  \frac{{{f_c}}}{c}\left[ {2{{\left\| {{\bf{x}}_k^v} \right\|}_2}\cos {\delta _k} \cos \left({\frac{{{\beta _k}}}{2}} \right) } \right. \\ & \left. { + {{\left\| {{\bf{x}}_{T,k}^v} \right\|}_2}\cos \left( {{\delta _{T,k}} - {\theta _{T,k}}} \right) + {{\left\| {{\bf{x}}_{R,k}^v} \right\|}_2}\cos \left( {{\delta _{R,k}} - {\theta _k}} \right)} \right]
\end{split}$}
\end{equation}
where ${\delta _k}$ is the angle measured from bisector vector ${\bf{v} _B}$ to target velocity vector ${\bf{x}}_k^v$ and restricted to $\left[ {0,2\pi } \right]$. ${\delta_{T,k}}$ and ${\delta_{R,k}}$ denote the angle of transmitter velocity vector ${\bf{x}}_{T,k}^v$ and receiver velocity vector ${\bf{x}}_{R,k}^v$, respectively,  both measured positive counter-clockwise from the x-axis and restricted to $\left[ {0,2\pi } \right]$.  ${\beta _k}$ is the bistatic angle between transmitter and receiver with vertex at the target and $\cos \left({\frac{{{\beta _k}}}{2}} \right)$  is calculated by 
\begin{equation}\label{eq:cos_beta/2}
\resizebox{.9\hsize}{!}{$
\cos \left( {\frac{{{\beta _k}}}{2}} \right){\rm{ = }}\left\{ \begin{array}{l}
\frac{{{d_k} + {L_k}\cos \left( {{\theta _k} - {\theta _{TR,k}}} \right)}}{{\sqrt {d_k^2 + L_k^2 + 2{d_k}{L_k}\cos \left( {{\theta _k} - {\theta _{TR,k}}} \right)} }}{\kern 3pt}, {\beta _k} \ne \pi {\kern 2pt} \\
{\kern 66pt} 0{\kern 50pt},otherwise
\end{array} \right.$}
\end{equation}
As shown in (\ref{eq:xi_dk}) and (\ref{eq:cos_beta/2}), the target Doppler frequency $\xi_k$ is collectively determined by T2R relative position through triangular geometry parameters $\left( {{d_k},{L_k},{\theta _k},{\theta _{TR,k}}} \right)$, and target velocity ${\bf{x}}_k^v$ ,  transmitter velocity ${\bf{x}}_{T,k}^v$, receiver velocity ${\bf{x}}_{R,k}^v$.

\subsection{Geometry-dependent target measurement uncertainty}\label{GeoDep-TMU}
At each observation time ${t_k}$, bistatic radar randomly returns a target-originated measurement ${{\bf{z}}_k}$ with detection probability less than one, and ${{\bf{z}}_k}$ is corrupted by statistical noise ${{\bf{w}}_k}$ that is usually modeled by white Gaussian with zero mean and covariance. Therefore, detection probability and error covariance collectively determine the TMU. More specifically, the higher the detection probability, the lower the uncertainty a radar returns a target measurement, while the smaller the error covariance, the less the uncertainty contained in target measurement. Reminder of this section concentrates on deriving the generalized form for TMU of bistatic radar, and giving an approximated solution to specify the impact of T2R geometry on measurement error covariance. 

To keep consistence with existing work, the target is assumed to follow a Swerling I model, thus its detection probability depends on the T2R geometry through the signal-to-noise ratio (SNR) and can be calculated by \cite{ref:18}
\begin{equation}\label{eq:Pd_xk}
    {P_d}\left( {{\Delta _k}} \right) = P_{{\rm{FA}}}^{1/\left( {1 + {\Psi _k}\left( {{\Delta _k}} \right)} \right)}
\end{equation}
where ${P_{FA}}$ is the false alarm rate and ${\Psi _k}\left( {{\Delta _k}} \right)$ is the SNR at the receiver. ${\Psi _k}\left( {{\Delta _k}} \right)$  is time-varying and depends on the T2R geometry ${\Delta _k}$ through the energy path-loss factor due to signal propagation, given by
\begin{equation} \label{eq:SNR}
\resizebox{.85\hsize}{!}{$
    {\Psi _k}\left( {{\Delta _k}} \right) = \frac{{{\vartheta _0}^4}}{{{{(\left\| {{\bf{x}}_k^p - {\bf{x}}_{T,k}^p} \right\|{_2}{\kern 1pt} \left\| {{\bf{x}}_k^p - {\bf{x}}_{R,k}^p} \right\|{_2})}^2}}} = \frac{{{\vartheta _0}^4}}{{{{({R_{T,k}}{R_{R,k}})}^2}}}$}
\end{equation}
where ${{\vartheta _0}}$ is the transmitted signal constant. As pointed out in \cite{ref:18}, the error of DOA is inversely proportional to the SNR ${\Psi _k}\left( {{\Delta _k}} \right)$ at the receiver and independent to the errors of bistatic range and velocity, therefore the error covariance of target bistatic radar measurement ${{\bf{z}}_k}$ can be approximated by its CRLB as
\begin{equation}\label{eq:Rk}
\begin{split}
{{\bf{R}}_k}\left( {{\Delta _k}} \right) & \approx {\rm{diag}}\left( {{\bf{C}}\left( {{d_k},{v_k}} \right),{\bf{C}}\left( {{\theta _k}} \right)} \right) \\ & = \left[ {\begin{array}{*{20}{c}}
{{{\left[ {{{\bf{J}}_M}({d_k},{v_k})} \right]}^{ - 1}}}&0\\ 
0&{\frac{{\sigma _{{\theta _0}}^2}}{{{\Psi _k}\left( {{\Delta _k}} \right)}}}
\end{array}} \right]
\end{split}
\end{equation}
where ${\sigma_{{\theta_0}}}$ is the standard deviation of referred DOA signal, ${\bf{C}}\left( {{d_k},{v_k}} \right)$ denotes the CRLB for bistatic range ${d_k}$ and bistatic velocity ${v_k}$, and ${\bf{C}}\left( {{\theta _k}} \right)$ denotes the CRLB for DOA ${\theta _k}$. ${{\bf{J}}_M}({d_k},{v_k})$ is the FIM for ${d_k}$ and ${v_k}$. Since the target bistatic measurement $\left( {{d_k}, {v_k}} \right)$ is converted from its bistatic signal $\left( {{\tau _k}, {\xi _k}} \right)$ via (\ref{eq:SigToMea}) and (\ref{eq:xi_dk}),  ${{\bf{J}}_M}({d_k},{v_k})$ can be therefore obtained by transforming the FIM for $\left( {{\tau _k}, {\xi _k}} \right)$ in signal domain to measurement domain as \cite{ref:18},\cite{ref:20},\cite{ref:33}
\begin{equation}\label{eq:FIM_JM}
{{\bf{J}}_{\rm{M}}}({d_k},{v_k}) = {{\bf{P}}_k}{{\bf{J}}_{\rm{S}}}(\tau_k ,\xi_k ){\bf{P}}_k^T
\end{equation}
where ${{\bf{J}}_S}({\tau _k},{\xi _k})$ denotes the FIM for ${\tau_k}$ and ${\xi_k}$, and is determined by both the SNR ${\Psi_k}\left({{\Delta_k}} \right)$ at the receiver and geometry-independent signal parameter \cite{ref:18}, \cite{ref:33},
\begin{equation} \label{eq:FIM_Js}
{{\bf{J}}_S}({\tau _k},{\xi _k}) = {\Psi _k}\left( {{\Delta _k}} \right)\left[ {\begin{array}{*{20}{c}}
{{S_1}}&{{S_2}}\\
{{S_2}}&{{S_3}}
\end{array}} \right]
\end{equation}
with ${S_1}$, ${S_3}$, ${S_3}$ denoting constants related to transmitted signal. ${{\bf{P}}_k}$ is the transformation matrix from the signal domain $\left( {{\tau _k},{\xi _k}} \right)$ to measurement domain $\left( {{d_k},{v_k}} \right)$, and depends on the states of both target and radar, given by
\begin{equation} \label{eq:P_SigToMea}
{{\bf{P}}_k} = \left[ {\begin{array}{*{20}{c}}
{\frac{{\partial {\tau _k}}}{{\partial {d_k}}}}&{\frac{{\partial {\xi _k}}}{{\partial {d_k}}}}\\
{\frac{{\partial {\tau _k}}}{{\partial {v_k}}}}&{\frac{{\partial {\xi _k}}}{{\partial {v_k}}}}
\end{array}} \right]
\end{equation}
with each element calculated as
\begin{equation}
    \frac{{\partial {\tau _k}}}{{\partial {d_k}}} = \frac{1}{c},{\rm{ }}\frac{{\partial {\tau _k}}}{{\partial {v_k}}} = 0,{\rm{ }}\frac{{\partial {\xi _k}}}{{\partial {v_k}}} = \frac{{{f_c}}}{c}
\end{equation}
in particular, $\frac{{\partial {\xi _k}}}{{\partial {d_k}}}$ is given by
\begin{equation}\label{eq:diff_xi_dk}
\resizebox{.87\hsize}{!}{$
\frac{{\partial {\xi _k}}}{{\partial {d_k}}} = \left\{ \begin{array}{l}
\frac{{{f_c}}}{c}\frac{{2{{\left\| {{\bf{x}}_k^v} \right\|}_2}\cos {\delta _k}L_k^2{{\sin }^2}\left( {{\theta _k} - {\theta _{TR,k}}} \right)}}{{\sqrt {{{\left( {d_k^2 + L_k^2 + 2{d_k}{L_k}\cos \left( {{\theta _k} - {\theta _{TR,k}}} \right)} \right)}^3}} }},{\beta _k} \ne \pi \\
{\kern 75pt} 0{\kern 55pt}, otherwise
\end{array} \right.$}
\end{equation}
\begin{figure*}[htb]
\hrulefill
\vspace*{2pt}
\centering
\begin{equation}\label{eq:Rk_Final}
\resizebox{.77\hsize}{!}{$
{{\bf{R}}_k}\left( {{\Delta _k}} \right) = \frac{1}{{{\Psi _k}\left( {{\Delta _k}} \right)}}\frac{{{c^4}}}{{f_c^2\left( {{S_1}{S_3} - S_2^2} \right)}}
\left[ \begin{array}{l}
\left[ \begin{array}{l}
{\kern 26pt} \frac{{f_c^2}}{{{c^2}}}{S_3}{\kern 62pt} - \frac{{{f_c}}}{{{c^2}}}{S_2} - \frac{{{f_c}}}{c}\frac{{\partial {\xi _k}}}{{\partial {d_k}}}{S_3}\\
 - \frac{{{f_c}}}{{{c^2}}}{S_2} - \frac{{{f_c}}}{c}\frac{{\partial {\xi _k}}}{{\partial {d_k}}}{S_3}{\kern 16pt} \frac{1}{{{c^2}}}{S_1} + \frac{2}{c}\frac{{\partial {\xi _k}}}{{\partial {d_k}}}{S_2} + {\left( {\frac{{\partial {\xi _k}}}{{\partial {d_k}}}} \right)^2}{S_3}
\end{array} \right]{\kern 23pt} 0\\
{\kern 106pt} 0{\kern 124pt} \frac{{\sigma _{{\theta _0}}^2f_c^2\left( {{S_1}{S_3} - S_2^2} \right)}}{{{c^4}}} 
\end{array} \right]$}
\end{equation}
\hrulefill
\vspace*{-2pt}
\end{figure*}

Substituting (\ref{eq:FIM_Js}-\ref{eq:P_SigToMea}) to (\ref{eq:FIM_JM}), whose result is then substituted to (\ref{eq:Rk}), the full expression of radar measurement error covariance can be finalized as shown in (\ref{eq:Rk_Final}). 

\textbf{Remark 1:} as seen in (\ref{eq:Rk_Final}), ${{\bf{R}}_k}\left( {{\Delta _k}} \right)$ is collectively determined by three types of factors, i.e., SNR ${\Psi _k}\left( {{\Delta _k}} \right)$, $\frac{{\partial {\xi _k}}}{{\partial {d_k}}}$ and signal parameters $\left( {{S_1},{S_2},{S_3}} \right)$.  Among these factors,  $\left( {{S_1},{S_2},{S_3}} \right)$ usually keeps fixed once transmitted signal is selected, while ${\Psi _k}\left( {{\Delta _k}} \right)$ depends on the T2R geometry ${\Delta _k}$ through term $R_{T,k}R_{R,k}$ shown in (\ref{eq:SNR}),  and $\frac{{\partial {\xi _k}}}{{\partial {d_k}}}$ relates to both T2R geometry ${\Delta_k}$ through $\left( {{d_k},{L_k},{\theta _k},{\theta _{TR,k}}} \right)$ and target velocity ${\bf{x}}_k^v$ as shown in (\ref{eq:diff_xi_dk}), either ${\Psi _k}\left( {{\Delta _k}} \right)$ or $\frac{{\partial {\xi _k}}}{{\partial {d_k}}}$ changes as any of the target and radar moves along its trajectory. Therefore, in realistic radar operation, target measurement error covariance ${{\bf{R}}_k}\left( {{\Delta _k}} \right)$ is time varying and collectively affected by both T2R geometry ${\Delta_k}$ and target velocity ${\bf{x}}_k^v$.

\textbf{Assumption 1}: to simplify (\ref{eq:Rk_Final}), we make an assumption $\frac{{\partial {\xi _k}}}{{\partial {d_k}}} \approx 0$ for any value of ${\beta _k} \ne \pi$, i.e., the target Doppler shift hardly changes as the bistatic range varies.  
\begin{figure}[htbp]
\centering
\subfloat[$\frac{{\partial {\xi _k}}}{{\partial {d_k}}}$ versus target DOA ${\theta _k}$]{\includegraphics[width=0.35\textwidth]{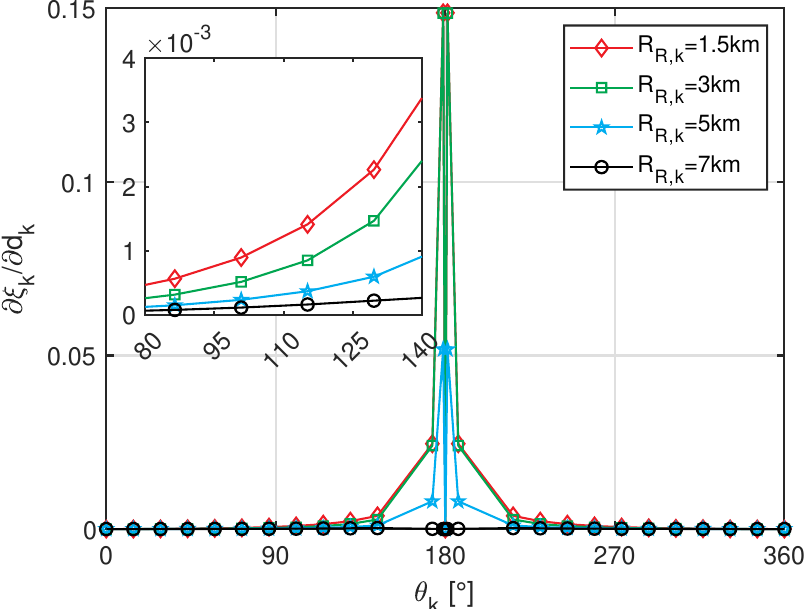}
\label{fig:Assum1}} \\
\subfloat[${\sigma_v}$ versus ${\theta_k}$ with and without assumption 1]{\includegraphics[width=0.35\textwidth]{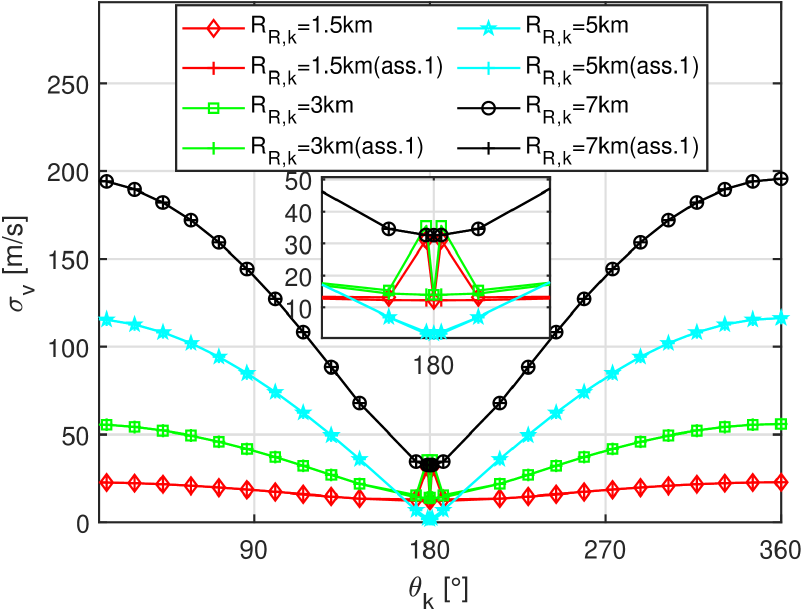}
\label{fig:SigmaV_Assum1}}
\caption{Numerical validation of feasibility of assumption 1}
\label{fig:Assump1Vad}
\end{figure}
The complicated expression in (\ref{eq:diff_xi_dk}) hampers us to mathematically analyze the approximation bias caused by Assumption 1. Instead, we investigate approximation bias via numerical experiments, in which T2R geometry is set to change as either ${R_{R,k}}$ or ${\theta _k}$ varies when target moves along its trajectory, with experimental setup detailed in section~\ref{sec:5}. As respectively shown  in Fig. \ref{fig:Assum1} and Fig. \ref{fig:SigmaV_Assum1}, as long as ${\theta _k} \neq {180}^o$, the value of $\frac{{\partial {\xi _k}}}{{\partial {d_k}}}$ approaches to zero and there is no nontrivial difference between ${\sigma_v}$ with and without assumption 1. Fortunately, the target is barely collinear with transmitter and receiver in realistic tracking scenario, hence approximation bias of assumption 1 can be acceptable. Consequently, ${{\bf{R}}_k}\left( {{\Delta _k}} \right)$ becomes only relevant to the T2R geometry ${\Delta _k}$ under assumption 1 and can be simplified as 
\begin{equation}\label{eq:Rk_xk}
\resizebox{.8\hsize}{!}{$
\begin{split}
{{\bf{R}}_k}\left( {{\Delta _k}} \right) \approx & \frac{1}{{\Psi \left( {{\Delta _k}} \right)}}\frac{{{c^4}}}{{f_c^2\left( {{S_1}{S_3} - S_2^2} \right)}} \\
& \times \left[ \begin{array}{l}
\left[ \begin{array}{l}
{\kern 5pt} \frac{{f_c^2}}{{{c^2}}}{S_3}{\kern 6pt}-\frac{{{f_c}}}{{{c^2}}}{S_2}\\
 - \frac{{{f_c}}}{{{c^2}}}{S_2}{\kern 16pt} \frac{1}{{{c^2}}}{S_1}
\end{array} \right]{\kern 37pt} 0\\
{\kern 32pt}  0{\kern 40pt} \frac{{\sigma _{{\theta _0}}^2f_c^2\left( {{S_1}{S_3} - S_2^2} \right)}}{{{c^4}}}
\end{array} \right]
\end{split}$}
\end{equation}

\section{Improved PCRLB fully accounting for radar measurement uncertainty}\label{sec:3}
In this section, we rigorously derive an improved  PCRLB (IPCRLB) for radar tracking in clutter by fully accounting for radar measurement uncertainty in terms of both MOU and geometry-dependent TMU. When deriving IPCRLB, the information gain factor (IGF) is specifically defined to quantify additional target Fisher information extracted from current measurements by considering the dependence of target detection probability and measurement error covariance on T2R geometry.  Since derivation of IPCRLB involves taking differentiating with respect to target state ${{\bf{x}}_k}$, for simplicity, all geometry-related parameters are abbreviated as the function of ${{\bf{x}}_k}$ at the rest of this paper, i.e., ${P_d}\left( {{\Delta _k}} \right) \buildrel \Delta \over = {P_d}\left( {{{\bf{x}}_k}} \right)$, ${{\bf{R}}_k}\left( {{\Delta _k}} \right) \buildrel \Delta \over = {{\bf{R}}_k}\left( {{{\bf{x}}_k}} \right)$ and ${\Psi _k}\left( {{\Delta _k}} \right) \buildrel \Delta \over = {\Psi _k}\left( {{{\bf{x}}_k}} \right)$.
\subsection{Derivation of IPCRLB}
Define ${\bf{Z}}(k) = \left\{ {{\bf{z}}_k^1, \cdots ,{\bf{z}}_k^{{m_k}}} \right\}$as the set of origin-unknown measurements returned by radar at time ${t_k}$,  ${\bf{z}}_k^i$ is the $i$-th measurement of ${\bf{Z}}(k)$, $i=1,...,m_k$, ${m_k}$ is the cardinality of ${\bf{Z}}(k)$. ${\bf{z}}_k^i$ can be originated from either target of interest or clutter. At each time ${t_k}$, the radar randomly returned the target-originated measurement with detection probability ${P_d}\left( {{{\bf{x}}_k}} \right)$, and returned target measurement is corrupted by white Gaussian noise with zero mean and covariance ${{\bf{R}}_k}\left( {{{\bf{x}}_k}} \right)$. Therefore, the likelihood of target measurement ${\bf{z}}_k^i$ conditioned on ${{\bf{x}}_k}$ can be described by a Gaussian
\begin{equation}
\begin{split}
& p({\bf{z}}_k^i)  \buildrel \Delta \over  =  p({\bf{z}}_k^i|{{\bf{x}}_k})\\
& {\rm{ = }}\frac{{\exp \left[ { - \frac{1}{2}{{\left( {{\bf{z}}_k^i - {\bf{h}}({{\bf{x}}_k})} \right)}^T}{\bf{R}}_k^{ - 1}({{\bf{x}}_k})\left( {{\bf{z}}_k^i - {\bf{h}}({{\bf{x}}_k})} \right)} \right]}}{{\sqrt {{{(2\pi )}^n}\left| {{{\bf{R}}_k}({{\bf{x}}_k})} \right|} }}
\end{split}
\end{equation}
where $n$ is the dimension of ${\bf{z}}_k^i$. The number of clutter measurements in ${\bf{Z}}(k)$ is assumed to follow a Poisson distribution with mean number $\lambda V$, with each clutter measurement is uniformly distributed over the radar measurement volume $V$ with spatial density $\lambda={\raise0.7ex\hbox{${{N_{cell}}{P_{FA}}}$}\mathord{\left/ {\vphantom {{{N_{cell}}{P_{FA}}} V}}\right.\kern-\nulldelimiterspace}\hbox{$V$}}$, ${N_{cell}}$ is the total number of radar resolution cells. Consequently, the probability that radar returns ${m_k}$ measurements at time ${t_k}$ can be calculated by \cite{ref:15}, \cite{ref:17}-\cite{ref:18}
\begin{equation}\label{eq:P_xkmk}
\small
\begin{split}
P({{\bf{x}}_k},{m_k})   = & \left[ {1 - {P_d}({{\bf{x}}_k})} \right]{e^{ - \lambda V}}\frac{{{{(\lambda V)}^{{m_k}}}}}{{{m_k}!}}  \\
& + {P_d}({{\bf{x}}_k}){e^{ - \lambda V}}\frac{{{{(\lambda V)}^{{m_k} - 1}}}}{{({m_k} - 1)!}}
\end{split}
\end{equation}
The probability that there is one measurement in ${\bf{Z}}(k)$ originated from target with state ${{\bf{x}}_k}$ can be described by \cite{ref:15}, \cite{ref:17}-\cite{ref:18}
\begin{equation}\label{eq:dk_xkmk}
d({{\bf{x}}_k},{m_k}) = \frac{{{P_d}({{\bf{x}}_k})}}{{\left( {1 - {P_d}({{\bf{x}}_k})} \right)\frac{{\lambda V}}{{{m_k}}} + {P_d}({{\bf{x}}_k})}}
\end{equation}
The likelihood of measurements ${\bf{Z}}(k)$ conditioned on target state ${{\bf{x}}_k}$ and number of measurements ${m_k}$ is given by \cite{ref:15}, \cite{ref:17}-\cite{ref:18}
\begin{equation}\label{eq:P_CondLikelihood}
\resizebox{.88\hsize}{!}{$
p({\bf{Z}}(k)|{{\bf{x}}_k},{m_k}) = \frac{{1 - d({{\bf{x}}_k},{m_k})}}{{{V^{{m_k}}}}} + \frac{{d({{\bf{x}}_k},{m_k})}}{{{m_k}{V^{{m_k} - 1}}}}\sum\limits_{i = 1}^{{m_k}} {p({\bf{z}}_k^i)}$}
\end{equation}
Let ${{\bf{J}}_k}$ denote the FIM for target state ${{\bf{x}}_k}$ at ${t_k}$, its inverse is defined as the PCRLB that bounds the MSE of any unbiased estimator \cite{ref:1}
\begin{equation}\label{eq:JK_Definition}
{\bf{J}}_k^{ - 1} \le {{\mathop{\rm E}\nolimits} _{{{\bf{x}}_k}}}\{ ({{\bf{\hat x}}_{k|k}} - {{\bf{x}}_k}){({{\bf{\hat x}}_{k|k}} - {{\bf{x}}_k})^T}\}
\end{equation}
where ${\mathop{\rm E}\nolimits} \{ .\}$ is the expectation operator. The Riccati-like recursion form to calculate ${{\bf{J}}_k}$ is given by  \cite{ref:2} 
\begin{equation}\label{eq:Jk_Recur}
{{\bf{J}}_k} = {({\bf{Q_k}} + {\bf{F}}_k {\bf J}_{k - 1}^{ - 1}{{\bf{F_k}}^T})^{ - 1}} + {\bf{J}}_k^{\bf{Z}}
\end{equation}
with ${\bf{F_k}}$ and ${\bf{Q_k}}$ denotes the target state propagation matrix and process noise covariance, respectively. The FIM in (\ref{eq:Jk_Recur}) consists of two parts: the target prior information up to time $t_{k-1}$ and the target new information contributed by measurement ${\bf{Z}}(k)$ received at $t_k$. The new information ${\bf{J}}_k^{\bf{Z}}$  is defined and calculated as \cite{ref:15}, \cite{ref:17}-\cite{ref:18}
\begin{equation}\label{eq:Jkz_Orig}
\begin{split}
{\bf{J}}_k^{\bf{Z}} & = {{\rm{E}}_{{{\bf{x}}_k}}}\left\{ {\left\{ {{{\rm{E}}_{{m_k}}}\{ {\bf{J}}_k^{\bf{Z}}({{\bf{x}}_k},{m_k})} \right\}} \right\} \\  
& = {{\rm{E}}_{{{\bf{x}}_k}}}\left\{ {\sum\limits_{{m_k} = 1}^\infty  {P({{\bf{x}}_k},{m_k})} {\bf{J}}_k^{\bf{Z}}({{\bf{x}}_k},{m_k})} \right\}
\end{split}
\end{equation}
where ${\bf{J}}_k^{\bf{Z}}({{\bf{x}}_k},{m_k})$ is the conditional measurement information matrix and defined as
\begin{equation}\label{eq:Jkz_Cond}
\small
\begin{split}
&{\bf{J}}_k^{\bf{Z}}({{\bf{x}}_k},{m_k}) \\
&= {{\rm{E}}_{\bf{Z}(k)}}\left\{ {\left[ {{\nabla _{{{\bf{x}}_k}}}\ln p({\bf{Z}}(k)|{{\bf{x}}_k},{m_k})} \right]{{\left[ {{\nabla _{{{\bf{x}}_k}}}\ln p({\bf{Z}}(k)|{{\bf{x}}_k},{m_k})} \right]}^T}} \right\}
\end{split}
\end{equation}
where ${\nabla _{{{\bf{x}}_k}}}$ denotes the first-order partial derivative operator with respect to ${{\bf{x}}_k}$. In particular, we treat both $P_D$ and $\bf{R}_k$ in $\ln p({\bf{Z}}(k)|{{\bf{x}}_k},{m_k})$ as target state ${{\bf{x}}_k}$ dependent parameters, and take the first-order partial derivative of the logarithm function of (\ref{eq:P_CondLikelihood}) w.r.t. $\bf{x}_k$. For simplicity, we follow \cite{ref:8,ref:9,ref:15, ref:18} to assume ${{\bf{R}}_k}({{\bf{x}}_k}) = diag\left( {\sigma _1^2, \cdots ,\sigma _n^2} \right)$, consequently, ${\nabla _{{{\bf{x}}_k}}}\ln p({\bf{Z}}(k)|{{\bf{x}}_k},{m_k})$ can be calculated and decomposed into three parts
\begin{equation}\label{eq:CondLike_diff}
{\nabla _{{{\bf{x}}_k}}}\ln p({\bf{Z}}(k)|{{\bf{x}}_k},{m_k}) = {\bf{A}} + {\bf{B}} + {\bf{C}}
\end{equation}
with $\bf{A}$, $\bf{B}$, $\bf{C}$ given by
\begin{equation}\label{eq:A}
\resizebox{.88\hsize}{!}{$
{\bf{A}} = \frac{{d({{\bf{x}}_k},{m_k}){\bf{H}}_k^T({{\bf{x}}_k}){\bf{R}}_k^{ - 1}({{\bf{x}}_k})}}{{{m_k}{V^{{m_k} - 1}}p({\bf{Z}}(k)|{{\bf{x}}_k},{m_k})}}\sum\limits_{i = 1}^{{m_k}} {p({\bf{z}}_k^i)\left[ {{\bf{z}}_k^i - {\bf{h}}({{\bf{x}}_k})} \right]}$}
\end{equation}
\begin{equation}\label{eq:B}
\resizebox{.88\hsize}{!}{$
{\bf{B}} = \frac{1}{{p({\bf{Z}}(k)|{{\bf{x}}_k},{m_k})}}\left[ {\frac{{\sum\limits_{i = 1}^{{m_k}} {p({\bf{z}}_k^i)} }}{{{m_k}{V^{{m_k} - 1}}}} - \frac{1}{{{V^{{m_k}}}}}} \right]\frac{{\partial d({{\bf{x}}_k},{m_k})}}{{\partial {{\bf{x}}_k}}}$}
\end{equation}
\begin{equation} \label{eq:C}
\resizebox{.88\hsize}{!}{$
\begin{split}
&{\bf{C}} =   \frac{-{d({{\bf{x}}_k},{m_k})}}{{2{\Psi _k}\left( {{{\bf{x}}_k}} \right){m_k}{V^{{m_k} - 1}}p({\bf{Z}}(k)|{{\bf{x}}_k},{m_k})}}\frac{{\partial {\Psi _k}\left( {{{\bf{x}}_k}} \right)}}{{\partial {{\bf{x}}_k}}} \times \\
& \sum\limits_{i = 1}^{{m_k}} {p({\bf{z}}_k^i)\left\{ {{{\left[ {{\bf{z}}_k^i - {\bf{h}}({{\bf{x}}_k})} \right]}^T}{\bf{R}}_k^{ - 1}({{\bf{x}}_k})\left[ {{\bf{z}}_k^i - {\bf{h}}({{\bf{x}}_k})} \right] - n} \right\}}
\end{split}$}
\end{equation}
where ${{\bf{H}}_k}({{\bf{x}}_k})$ is the Jacobian matrix of measurement function ${\bf{h}}({{\bf{x}}_k})$. Please note, compared to existing work that treats both $P_d$ and $\bf{R}_k$ as state-independent parameters when taking the partial derivative of $\ln p({\bf{Z}}(k)|{{\bf{x}}_k},{m_k})$ w.r.t $\bf{x}_k$, the result in (\ref{eq:CondLike_diff} ) contains two more terms, i.e.,  $\bf{B}$ and $\bf{C}$. In particular, $\bf{B}$ results from treating $P_d$ as state-dependent parameter while $\bf{C}$ results from treating $\bf{R}_k$ as state-dependent parameter when differentiating $\ln p({\bf{Z}}(k)|{{\bf{x}}_k},{m_k})$ w.r.t $\bf{x}_k$. Taking the first-order partial derivative of (\ref{eq:dk_xkmk}) and (\ref{eq:SNR}) w.r.s. to $\bf{x}_k$, respectively,  we have
\begin{equation}
\begin{split}
\frac{{\partial d({{\bf{x}}_k},{m_k})}}{{\partial {{\bf{x}}_k}}} =   & \frac{{- \lambda V{m_k}P_{{\rm{FA}}}^{1/\left( {1 + {\Psi _k}\left( {{{\bf{x}}_k}} \right) } \right)}}}{{{{\left\{ {\left[ {1 - {P_d}({{\bf{x}}_k})} \right]\lambda V + {m_k}{P_d}({{\bf{x}}_k})} \right\}}^2}}}  \\
& \times \frac{{\ln {P_{{\rm{FA}}}}}}{{{{\left[ {1 + {\Psi _k}\left( {{{\bf{x}}_k}} \right)} \right]}^2}}}\frac{{\partial {\Psi _k}\left( {{{\bf{x}}_k}} \right)}}{{\partial {{\bf{x}}_k}}}
\end{split}
\end{equation}
\begin{equation}\label{eq:SNR_diff}
\resizebox{.88\hsize}{!}{$
\begin{split}
&\frac{{\partial {\Psi _k}\left( {{{\bf{x}}_k}} \right)}}{{\partial {{\bf{x}}_k}}} = \frac{{ - 2R_0^4}}{{R_{R,k}^4R_{T,k}^4}}  \\
& \times {\left[ {{\kern 1pt} \left( {\left[ {\left( {{\bf{x}}_k^p - {\bf{x}}_{R,k}^p} \right){\kern 5pt} \left( {{\bf{x}}_k^p - {\bf{x}}_{T,k}^p} \right)} \right]\left[ \begin{array}{l}
R_{T,k}^2\\
R_{R,k}^2{\kern 1pt} 
\end{array} \right]} \right)^T{\kern 12pt} {{\bf{0}}_{1 \times 2}}} \right]^T}
\end{split}$}
\end{equation}
By substituting (\ref{eq:CondLike_diff}) to (\ref{eq:Jkz_Cond}),  ${\bf{J}}_k^{\bf{Z}}({{\bf{x}}_k},{m_k})$ can be then written as (\ref{eq:Jkz_xkmkOrg}). 
\begin{figure*}[htb]
\centering
\hrulefill
\vspace*{2pt}
\begin{equation}\label{eq:Jkz_xkmkOrg}
\resizebox{.85\hsize}{!}{$
\begin{split}
&{\bf{J}}_k^{\bf{Z}}({{\bf{x}}_k},{m_k}) = {{\rm{E}}_{{{\bf{Z}}(k)}}}\left\{ ({\bf{A}}{{\bf{A}}^T} + {\bf{A}}{{\bf{B}}^T} + {\bf{A}}{{\bf{C}}^T} + {\bf{B}}{{\bf{A}}^T} + {\bf{B}}{{\bf{B}}^T} + {\bf{B}}{{\bf{C}}^T} + {\bf{C}}{{\bf{A}}^T} + {\bf{C}}{{\bf{B}}^T} + {\bf{C}}{{\bf{C}}^T}) \right\}\\
& = {\int\limits_{{\bf{z}}_k^1 \in \Xi} { \cdots \int\limits_{{\bf{z}}_k^{{m_k}} \in \Xi} {({\bf{A}}{{\bf{A}}^T} + {\bf{A}}{{\bf{B}}^T} + {\bf{A}}{{\bf{C}}^T} + {\bf{B}}{{\bf{A}}^T} + {\bf{B}}{{\bf{B}}^T} + {\bf{B}}{{\bf{C}}^T} + {\bf{C}}{{\bf{A}}^T} + {\bf{C}}{{\bf{B}}^T} + {\bf{C}}{{\bf{C}}^T})p({\bf{Z}}(k)|{{\bf{x}}_k},{m_k})} d} {\bf{z}}_k^1 \cdots {\bf{z}}_k^{{m_k}}}
\end{split}$}
\end{equation}
\hrulefill
\vspace*{-2pt}
\end{figure*}
${\Xi}$ in (\ref{eq:Jkz_xkmkOrg}) denotes the integration region of ${\bf{z}}_k^i, i = 1, \ldots ,{m_k}$,  the volume of ${\Xi}$ is $V$. If we make the following change of variable:
\begin{equation}
{\bf{\tilde z}}_k^i = {\bf{z}}_k^i - {\bf{h}}({{\bf{x}}_k}), {\kern 5pt}{\bf{\tilde z}}_k^i \in \tilde \Xi
\end{equation}
$\tilde \Xi$ is the mapping of $\Xi$ and becomes symmetric domain under the above transformation, the volume of $\tilde \Xi$ is still $V$, let ${\bf{\tilde Z}}(k) = \left\{ {{\bf{\tilde z}}_k^1, \cdots,{\bf{\tilde z}}_k^{{m_k}}} \right\}$. Since ${\bf{\tilde z}}_k^i$ is an odd symmetric function of itself, while ${{\left[ {\bf{\tilde z}}_k^i \right]}^T}{\bf{R}}_k^{ - 1}({{\bf{x}}_k}) {\bf{\tilde z}}_k^i$ is an even symmetric function of ${\bf{\tilde z}}_k^i$,  therefore, ${\bf{A}}$ is an odd symmetric function of ${\bf{\tilde z}}_k^i$,  whereas, both ${\bf{B}}$, ${\bf{C}}$ and $p({\bf{\tilde Z}}(k)|{{\bf{x}}_k},{m_k})$ become even symmetric functions of ${\bf{\tilde z}}_k^i$ for $i=1,...,m_k$. Integrating an odd symmetric function over a symmetric domain $\tilde \Xi$, we have 
\begin{equation}
\resizebox{.88\hsize}{!}{$
\begin{split}
\left\{ \begin{array}{l}
\int\limits_{{\bf{\tilde z}}_k^1 \in \tilde \Xi} { \cdots \int\limits_{{\bf{\tilde z}}_k^{m_k} \in \tilde \Xi} {{\bf{A}}{{\bf{B}}^T}p({\bf{\tilde Z}}(k)|{{\bf{x}}_k},{m_k})} d} {\bf{\tilde z}}_k^1 \cdots d{\bf{\tilde z}}_k^{{m_k}} = 0\\
\int\limits_{{\bf{\tilde z}}_k^1 \in \tilde \Xi} { \cdots \int\limits_{{\bf{\tilde z}}_k^{m_k} \in \tilde \Xi} {{\bf{B}}{{\bf{A}}^T}p({\bf{Z}}(k)|{{\bf{x}}_k},{m_k})} d} {\bf{\tilde z}}_k^1 \cdots d{\bf{\tilde z}}_k^{{m_k}} = 0\\
\int\limits_{{\bf{\tilde z}}_k^1 \in \tilde \Xi} { \cdots \int\limits_{{\bf{\tilde z}}_k^{m_k} \in \tilde \Xi} {{\bf{A}}{{\bf{C}}^T}p({\bf{\tilde Z}}(k)|{{\bf{x}}_k},{m_k})} d} {\bf{\tilde z}}_k^1 \cdots d{\bf{\tilde z}}_k^{{m_k}} = 0\\
\int\limits_{{\bf{\tilde z}}_k^1 \in \tilde \Xi} { \cdots \int\limits_{{\bf{\tilde z}}_k^{m_k} \in \tilde \Xi} {{\bf{C}}{{\bf{A}}^T}p({\bf{\tilde Z}}(k)|{{\bf{x}}_k},{m_k})} d} {\bf{\tilde z}}_k^1 \cdots d{\bf{\tilde z}}_k^{{m_k}} = 0
\end{array} \right.
\end{split}$}
\end{equation}

Hence,  based on the above results, along with the fact ${\bf{B}}{{\bf{C}}^T}={\bf{C}}{{\bf{B}}^T}$, (\ref{eq:Jkz_xkmkOrg}) can be simplified as (\ref{eq:Jkz_xkmkSimp}).
\begin{figure*}[htb]
\centering
\begin{equation}\label{eq:Jkz_xkmkSimp}
\resizebox{.75\hsize}{!}{$
{\bf{J}}_k^{\bf{Z}}({{\bf{x}}_k},{m_k}) = \int\limits_{{\bf{\tilde z}}_k^1 \in \tilde \Xi} { \cdots \int\limits_{{\bf{\tilde z}}_k^{m_k} \in \tilde \Xi} {\left\{ {{\bf{A}}{{\bf{A}}^T}+ {\bf{B}}{{\bf{B}}^T} + 2{\bf{B}}{{\bf{C}}^T} + {\bf{C}}{{\bf{C}}^T}} \right\}p({\bf{\tilde Z}}(k)|{{\bf{x}}_k},{m_k})} d{\bf{\tilde z}}_k^1 \cdots d{\bf{\tilde z}}_k^{{m_k}}} $}
\end{equation}
\hrulefill
\vspace*{2pt}
\end{figure*}
Then substitute (\ref{eq:A}) - (\ref{eq:C}) to (\ref{eq:Jkz_xkmkSimp}),  we further have
\begin{equation}\label{eq:Jkz_xkmkGen}
\begin{split}
{\bf{J}}_k^{\bf{Z}}({{\bf{x}}_k},{m_k}) & = \underbrace {{\bf{H}}_k^T({{\bf{x}}_k}){\bf{\Lambda }}_k^1({{\bf{x}}_k},{m_k}){\bf{R}}_k^{ - 1}({{\bf{x}}_k}){{\bf{H}}_k}({{\bf{x}}_k})}_{{{\rm{E}}_{{\bf{\tilde Z}}\left( k \right)}}\left\{ {{\bf{A}}{{\bf{A}}^T}} \right\}} \\
& + \underbrace {\Lambda _k^2({{\bf{x}}_k},{m_k})\frac{{\partial {\Psi _k}({{\bf{x}}_k})}}{{\partial {{\bf{x}}_k}}}{{\left[ {\frac{{\partial {\Psi _k}({{\bf{x}}_k})}}{{\partial {{\bf{x}}_k}}}} \right]}^T}}_{{{\rm{E}}_{{\bf{\tilde Z}}\left( k \right)}}\left\{ {{\bf{B}}{{\bf{B}}^T} + 2{\bf{B}}{{\bf{C}}^T} + {\bf{C}}{{\bf{C}}^T}} \right\}}
\end{split}
\end{equation}
where matrix ${\bf{\Lambda }}_k^1({{\bf{x}}_k},{m_k})$ is given by
\begin{equation}\label{eq:Lambda_k1Orig}
\small
\begin{split}
{\bf{\Lambda }}_k^1({{\bf{x}}_k},{m_k}) = & {{\rm{E}}_{{\bf{\tilde Z}}(k)}}\left\{ {\frac{{{d^2}({{\bf{x}}_k},{m_k}) {\bf{R}}_k^{ - 1}({{\bf{x}}_k})}}{{m_k^2{V^{2{m_k} - 2}}{p^2}({\bf{\tilde Z}}(k)|{{\bf{x}}_k},{m_k})}}} \right. \\
& \left.  { \times \sum\limits_{i = 1}^{{m_k}} {\sum\limits_{j = 1}^{{m_k}} {{p}({\bf{\tilde z}}_k^i){\bf{\tilde z}}_k^i{p}({\bf{\tilde z}}_k^j){{({\bf{\tilde z}}_k^j)}^T}} } } \right\}
\end{split}
\end{equation}
and scalar ${{\Lambda }}_k^2({{\bf{x}}_k},{m_k})$ is defined by
\begin{equation}\label{eq:Lambda_k2Orig}
\small
{{\Lambda }}_k^2({{\bf{x}}_k},{m_k}) = {{\rm{E}}_{{\bf{\tilde Z}}(k)}}\left\{ {\left( {{\Gamma _{\bf{B}\bf{B}^T}}+ {\Gamma _{\bf{B}\bf{C}^T}}+{\Gamma _{\bf{C}\bf{C}^T}}} \right)} \right\}
\end{equation}
with
\begin{equation}
\small
\begin{split}
& {\Gamma _{\bf{B}\bf{B}^T}}   = \frac{{{\lambda ^2}{V^2}m_k^2P_{{\rm{FA}}}^{2/\left( {1 + \Psi_k ({{\bf{x}}_k})} \right)}}}{{{{\left\{ {\left[ {1 - {P_d}({{\bf{x}}_k})} \right]\lambda V + {m_k}{P_d}({{\bf{x}}_k})} \right\}}^4}}}\\
& \times \frac{{{{\ln }^2}{P_{{\rm{FA}}}}}}{{{{[1 + \Psi_k ({{\bf{x}}_k})]}^4}}}\frac{1}{{{p^2}({\bf{\tilde Z}}(k)|{{\bf{x}}_k},{m_k})}}  \\
& \times \left[ {\frac{{\sum\limits_{i = 1}^{{m_k}} {{p}({\bf{\tilde z}}_k^i)} }}{{{m_k}{V^{{m_k} - 1}}}} - \frac{1}{{{V^{{m_k}}}}}} \right]\left[ {\frac{{\sum\limits_{j = 1}^{{m_k}} {{p}({\bf{\tilde z}}_k^j)} }}{{{m_k}{V^{{m_k} - 1}}}} - \frac{1}{{{V^{{m_k}}}}}} \right]
\end{split}
\end{equation}
\begin{equation}
\small
\resizebox{.88\hsize}{!}{$
\begin{split}
& {\Gamma _{\bf{B}\bf{C}^T}}  = \frac{{d({{\bf{x}}_k},{m_k})}}{{\Psi_k ({{\bf{x}}_k}){V^{{m_k} - 2}}}}\frac{{\ln {P_{{\rm{FA}}}}}}{{{{[1 + \Psi_k ({{\bf{x}}_k})]}^2}}} \\
& \times \frac{{\lambda P_{{\rm{FA}}}^{1/\left( {1 + \Psi_k ({{\bf{x}}_k})} \right)}}}{{{{\left\{ {\left[ {1 - {P_d}({{\bf{x}}_k})} \right]\lambda V + {m_k}{P_d}({{\bf{x}}_k})} \right\}}^2}}} \frac{1}{{{p^2}({\bf{\tilde Z}}(k)|{{\bf{x}}_k},{m_k})}} \\
& \times \left[ {\frac{{\sum\limits_{i = 1}^{{m_k}} {{p}({\bf{\tilde z}}_k^i)} }}{{{m_k}{V^{{m_k} - 1}}}} - \frac{1}{{{V^{{m_k}}}}}} \right]\sum\limits_{j = 1}^{{m_k}} {{p}({\bf{\tilde z}}_k^j)\left\{ {{{({\bf{\tilde z}}_k^j)}^T}{\bf{R}}_k^{ - 1}({{\bf{x}}_k}){\bf{\tilde z}}_k^j -n} \right\}}
\end{split}$}
\end{equation}
\begin{equation}
\small
\begin{split}
& {\Gamma _{\bf{C}\bf{C}^T}}  = \\
 & \frac{{{d^2}({{\bf{x}}_k},{m_k})}}{{4{\Psi ^2}({{\bf{x}}_k})m_k^2{V^{2{m_k} - 2}}}}\sum\limits_{i = 1}^{{m_k}} {{p}({\bf{\tilde z}}_k^i)\left\{ {{{({\bf{\tilde z}}_k^i)}^T}{\bf{R}}_k^{ - 1}({{\bf{x}}_k}){\bf{\tilde z}}_k^i -n} \right\}} \\
& \times \frac{1}{{{p^2}({\bf{\tilde Z}}(k)|{{\bf{x}}_k},{m_k})}}\sum\limits_{j = 1}^{{m_k}} {{p}({\bf{\tilde z}}_k^j)\left\{ {{{({\bf{\tilde z}}_k^j)}^T}{\bf{R}}_k^{ - 1}({{\bf{x}}_k}){\bf{\tilde z}}_k^j -n} \right\}}
\end{split}
\end{equation}

Note that, the first term on the right hand side of  ${\bf{J}}_k^{\bf{Z}}({{\bf{x}}_k},{m_k})$ shown in (\ref{eq:Jkz_xkmkGen}) is obtained by taking expectation of ${\bf{A}}{{\bf{A}}^T}$ w.r.t. ${\bf{\tilde Z}}(k)$, whose form is exactly the same as \cite{ref:8,ref:9,ref:15, ref:18} that only consider the MOU and totally ignore the geometry-dependent TMU when differentiating $\ln p({\bf{Z}}(k)|{{\bf{x}}_k},{m_k})$ w.r.t $\bf{x}_k$.  While, the second term in (\ref{eq:Jkz_xkmkGen}) is obtained by taking expectation of ${{\bf{B}}{{\bf{B}}^T} + 2{\bf{B}}{{\bf{C}}^T} + {\bf{C}}{{\bf{C}}^T}}$ w.r.t. ${\bf{\tilde Z}}(k)$, and acts as the additional conditional measurement information matrix stemmed from specifically treating both $P_d$ and $\bf{R}_k$ state-dependent parameters when differentiating $\ln p({\bf{Z}}(k)|{{\bf{x}}_k},{m_k})$ w.r.t $\bf{x}_k$.
To further calculate the conditional measurement information matrix derived in (\ref{eq:Jkz_xkmkGen}), one needs to deal with complicated expectations w.r.t. $\bf{\tilde Z}(k)$ appeared in ( \ref{eq:Lambda_k1Orig}) and (\ref{eq:Lambda_k2Orig}). To keep consistence with \cite{ref:8,ref:9,ref:15, ref:18}, the followed assumption is made to simplify the calculation of ( \ref{eq:Lambda_k1Orig}) and (\ref{eq:Lambda_k2Orig}) .

\textbf{Assumption 2}:  since the "far out" measurements will be given very small weight when using a reasonable data association algorithm, measurements are restricted to a validation gate of size $g$ times standard deviations, i.e.,
\begin{equation}\label{eq:zki_tilde}
\left| {{\bf{\tilde z}}_k^i[h]} \right| = |{\bf{z}}_k^i[h] - {{\bf{h}}^h}({{\bf{x}}_k})| < g{\sigma _h},{\kern 1pt} {\kern 1pt} {\kern 1pt} {\kern 1pt} h = 1,...,n
\end{equation}
with ${\bf{\tilde z}}_k^i = {\left[ {{\bf{\tilde z}}_k^i\left[ 1 \right]{\kern 1pt} {\kern 1pt} {\kern 1pt}  \cdots {\kern 1pt} {\kern 1pt} {\kern 1pt} {\kern 1pt} {\bf{\tilde z}}_k^i\left[ n \right]} \right]^T}$, ${\bf{\tilde z}}_k^i[h]$ denotes the $h$-th component of measurement ${\bf{\tilde z}}_k^i$, and $\sigma_h$ is the standard deviation of $h$-th component of ${\bf{\tilde z}}_k^i$. In the bistatic radar system introduced in section \ref{sec:2}, $n =3$, corresponding to the bistatic range, bistatic velocity and DOA, respectively. 

\textbf{Assumption 3}: the measurement dimensions are assumed to be orthogonal. This assumption gives good approximation for small integral regions,  which can be described by $n$-dimensional cube:
\begin{equation}\label{eq:Hat_Xi}
{\rm{\hat \Xi}} = {\rm{[ - g}}{\sigma _1}{\rm{,g}}{\sigma _1}{\rm{]}} \times  \cdots  \times {\rm{[ - g}}{\sigma _n}{\rm{,g}}{\sigma _n}{\rm{]}}
\end{equation}
with its corresponding volume:
\begin{equation}
{{\rm{V}}_g} = {(2g)^n}\mathop \prod \limits_{h = 1}^n {\sigma _h}
\end{equation}
Note that the typical value of $g$ is usually set to be 3 or 4, which means the probability that a target-generated measurement falling inside the validation gate is nearly one, therefore, the target detection probability inside the gate denoted by ${P^g_d}({{\bf{x}}_k})$ is almost equal to target detection probability ${P_d}({{\bf{x}}_k})$,  i.e., ${P_d^g({{\bf{x}}_k}) \approx {P_d}({{\bf{x}}_k})}$ . Similarly, other parameters can also be reasonably approximated inside gate, i.e., ${P^g_{{\rm{FA}}}} \approx {P_{{\rm{FA}}}}$, $P_g({{\bf{x}}_k},{m_k})\approx P({{\bf{x}}_k},{m_k})$, and $d_g({{\bf{x}}_k},{m_k}) \approx d({{\bf{x}}_k},{m_k})$.  Based upon these considerations, we derive  ${{\Lambda }}_k^2({{\bf{x}}_k},{m_k})$ in Appendix B in supplementary material, and its final result is given in (\ref{eq:Lambda_k2Final}),
\begin{figure*}[htb]
\hrulefill
\vspace*{2pt}
\centering
\begin{equation}\label{eq:Lambda_k2Final}
\resizebox{.75\hsize}{!}{$
\Lambda _k^2({{\bf{x}}_k},{m_k}) = {\left| {{{\bf{R}}_k}({{\bf{x}}_k})} \right|^{{m_k}/2}} \int\limits_{ - g}^g  \cdots  \int\limits_{ - g}^g  \cdots  \int\limits_{ - g}^g  \cdots  \int\limits_{ - g}^g {\frac{{\left( {{\Upsilon _1} + {\Upsilon _2} + {\Upsilon _3}} \right)}}{{\beta ({{\bf{x}}_k},{m_k})}}{\rm{d}}{\bf{\hat z}}_k^1\left[ 1 \right] \cdots {\rm{d}}{\bf{\hat z}}_k^1\left[ n \right] \cdots } {\rm{d}}{\bf{\hat z}}_k^{{m_k}}\left[ 1 \right] \cdots {\rm{d}}{\bf{\hat z}}_k^{{m_k}}\left[ n \right]$}
\end{equation}
\end{figure*}
in which ${\bf{\hat z}}_k^i\left[ h \right] = {\bf{\tilde z}}_k^i\left[ h \right]/{\sigma _h},h = 1,...,n$, and  parameters ${\Upsilon _1}$, ${\Upsilon _2}$ and ${\Upsilon _3}$ are calculated by (\ref{eq:Upsilon_1})-(\ref{eq:Upsilon_3}), respectively.
\begin{figure*}[htb]
\small
\hrulefill
\vspace*{2pt}
\centering
\begin{equation}\label{eq:Upsilon_1}
\resizebox{.95\hsize}{!}{$
{\Upsilon _1} = \frac{{{\lambda ^2}V_g^2m_k^2{{(P_{{\rm{FA}}}^g)}^{2/\left( {1 + {\Psi _k}({{\bf{x}}_k})} \right)}}{{\ln }^2}P_{{\rm{FA}}}^g}}{{\ell _g^4{{[1 + {\Psi _k}({{\bf{x}}_k})]}^4}}}\left\{ {\frac{{{e^{ - \sum\limits_{h = 1}^n {{{\left( {{\bf{\hat z}}_k^1[h]} \right)}^2}} }} + ({m_k} - 1){e^{ - \frac{{\sum\limits_{h = 1}^n {{{\left( {{\bf{\hat z}}_k^1[h]} \right)}^2} + {{\left( {{\bf{\hat z}}_k^2[h]} \right)}^2}} }}{2}}}}}{{{m_k}V_g^{2{m_k} - 2}{{\left( {2\pi } \right)}^n}\left| {{{\bf{R}}_k}\left( {{{\bf{x}}_k}} \right)} \right|}} - \frac{{2{e^{ - \frac{{\sum\limits_{h = 1}^n {{{\left( {{\bf{\hat z}}_k^1[h]} \right)}^2}} }}{2}}}}}{{V_g^{2{m_k} - 1}\sqrt {{{\left( {2\pi } \right)}^n}\left| {{{\bf{R}}_k}\left( {{{\bf{x}}_k}} \right)} \right|} }} + \frac{1}{{V_g^{2{m_k}}}}} \right\}$}
\end{equation}
\end{figure*}
\begin{figure*}[htb]
\small
\hrulefill
\vspace*{2pt}
\centering
\begin{equation}\label{eq:Upsilon_2}
\small
\resizebox{.95\hsize}{!}{$
\begin{split}
& {\Upsilon _2}  = \frac{{ {d_g}({{\bf{x}}_k},{m_k})}}{{{\Psi _k}({{\bf{x}}_k})V_g^{{m_k} - 2}}}\frac{{\lambda {{(P_{{\rm{FA}}}^g)}^{1/\left( {1 + {\Psi _k}({{\bf{x}}_k})} \right)}}\ln P_{{\rm{FA}}}^g}}{{\ell _g^2{{[1 + {\Psi _k}({{\bf{x}}_k})]}^2}}} \times \\
& \left\{ {\frac{{{e^{ - \sum\limits_{h = 1}^n {{{\left( {{\bf{\hat z}}_k^1[h]} \right)}^2}} }}\left[ {\sum\limits_{h = 1}^n {{{\left( {{\bf{\hat z}}_k^1[h]} \right)}^2}}  - n} \right] + \left( {{m_k} - 1} \right){e^{ - \frac{{\sum\limits_{h = 1}^n {{{\left( {{\bf{\hat z}}_k^1[h]} \right)}^2}}  + \sum\limits_{h = 1}^n {{{\left( {{\bf{\hat z}}_k^2[h]} \right)}^2}} }}{2}}}\left[ {\sum\limits_{h = 1}^n {{{\left( {{\bf{\hat z}}_k^2[h]} \right)}^2}}  - n} \right]}}{{V_g^{{m_k} - 1}{{\left( {2\pi } \right)}^n}\left| {{{\bf{R}}_k}\left( {{{\bf{x}}_k}} \right)} \right|}} - \frac{{{m_k}{e^{ - \frac{{\sum\limits_{h = 1}^n {{{\left( {{\bf{\hat z}}_k^1[h]} \right)}^2}} }}{2}}}\left[ {\sum\limits_{h = 1}^n {{{\left( {{\bf{\hat z}}_k^1[h]} \right)}^2}}  - n} \right]}}{{V_g^{{m_k}}\sqrt {{{\left( {2\pi } \right)}^n}\left| {{{\bf{R}}_k}\left( {{{\bf{x}}_k}} \right)} \right|} }}} \right\}
\end{split}$}
\end{equation}
\hrulefill
\vspace*{-2pt}
\end{figure*}
\begin{figure*}[htb]
\vspace*{-2pt}
\small
\centering
\begin{equation}\label{eq:Upsilon_3}
\small
\resizebox{.87\hsize}{!}{$
\begin{split}
{\Upsilon _3} & = \frac{{d_g^2({{\bf{x}}_k},{m_k})}}{{4\Psi _k^2({{\bf{x}}_k})m_k^2V_g^{2{m_k} - 2}{{\left( {2\pi } \right)}^n}\left| {{{\bf{R}}_k}\left( {{{\bf{x}}_k}} \right)} \right|}} \times \\
 & \left\{ {{m_k}\left( {{m_k} - 1} \right){e^{ - \frac{{\sum\limits_{h = 1}^n {{{\left( {{\bf{\hat z}}_k^1[h]} \right)}^2} + {{\left( {{\bf{\hat z}}_k^2[h]} \right)}^2}} }}{2}}}\left[ {\sum\limits_{h = 1}^n {{{\left( {{\bf{\hat z}}_k^1[h]} \right)}^2}}  - n} \right]\left[ {\sum\limits_{h = 1}^n {{{\left( {{\bf{\hat z}}_k^2[h]} \right)}^2}}  - n} \right] + {m_k}{e^{ - \sum\limits_{h = 1}^n {{{\left( {{\bf{\hat z}}_k^1[h]} \right)}^2}} }}{{\left[ {\sum\limits_{h = 1}^n {{{\left( {{\bf{\hat z}}_k^1[h]} \right)}^2}}  - n} \right]}^2}} \right\}
\end{split}$}
\end{equation}
\hrulefill
\end{figure*}
Since ${\bf{\Lambda }}_k^1({{\bf{x}}_k},{m_k})$ is exactly the same information reduction factor (IRF) as existing bound, we directly use the final equation from \cite{ref:8,ref:9,ref:15, ref:18}:
\begin{equation}
 {\bf{\Lambda }}_k^1({{\bf{x}}_k},{m_k}) = \Lambda _k^1({{\bf{x}}_k},{m_k}) \times {{\bf{I}}_n}   
\end{equation}
where ${{\bf{I}}_n}$ is the identity matrix with dimension $n$, $\Lambda _k^1({{\bf{x}}_k},{m_k})$ is calculated by (\ref{eq:Lambda_k1mkFinal}),
\begin{figure*}[htb]
\small
\hrulefill
\vspace*{2pt}
\centering
\begin{equation}\label{eq:Lambda_k1mkFinal}
\resizebox{.95\hsize}{!}{$
\begin{split}
\Lambda _k^1({{\bf{x}}_k},{m_k}) = \frac{{d_g^2({{\bf{x}}_k},{m_k}){{\left| {{{\bf{R}}_k}({{\bf{x}}_k})} \right|}^{({m_k} - 2)/2}}}}{{{m_k}V_g^{2{m_k} - 2}{{(2\pi )}^n}}} \int\limits_{ - g}^g  \cdots  \int\limits_{ - g}^g  \cdots  \int\limits_{ - g}^g  \cdots  \int\limits_{ - g}^g {\frac{{{{\left( {{\bf{\hat z}}_k^1\left[ 1 \right]} \right)}^2}{{e^{ - \sum\limits_{h = 1}^n {{{\left( {{\bf{\hat z}}_k^1\left[ h \right]} \right)}^2}} }}}}}{{\beta ({{\bf{x}}_k},{m_k})}}{\rm{d}}{\bf{\hat z}}_k^1\left[ 1 \right] \cdots {\rm{d}}{\bf{\hat z}}_k^1\left[ n \right] \cdots } {\rm{d}}{\bf{\hat z}}_k^{{m_k}}\left[ 1 \right] \cdots {\rm{d}}{\bf{\hat z}}_k^{{m_k}}\left[ n \right]
\end{split}$}
\end{equation}
\end{figure*}
 with $\beta ({{\bf{x}}_k},{m_k})$ given by
\begin{equation}
\begin{split}
\beta ({{\bf{x}}_k},{m_k}) & = \frac{{1 - {d_g}({{\bf{x}}_k},{m_k})}}{{V_g^{{m_k}}}} \\
& + \frac{{{d_g}({{\bf{x}}_k},{m_k})}}{{V_g^{{m_k} - 1}\sqrt {{{(2\pi )}^n}\left| {{{\bf{R}}_k}({{\bf{x}}_k})} \right|} }}{e^{ - \frac{{\sum\limits_{h = 1}^n {{{\left( {{\bf{\hat z}}_k^1\left[ h \right]} \right)}^2}} }}{2}}}
\end{split}
\end{equation}
Note that, to solve the integral involved in  (\ref{eq:Lambda_k2Final}) and (\ref{eq:Lambda_k1mkFinal}), we follow \cite{ref:8,ref:9,ref:15, ref:18}, and utilize the principle of importance sampling to carry out a Monte Carlo approximation, i.e., drawing sufficient realizations of each independently and identically distributed random variable ${\bf{\hat z}}_k^{{i}}\left[ n \right]$ from a uniform distribution in $\left[ { - g,{\kern 1pt} g} \right]$.  Furthermore, the sum on the right hand side of (\ref{eq:Jkz_Orig}) goes from one to infinity, for simplicity, the sum can be truncated to a small $m_k$, such as, $m_k = 2$ or $3$, its approximation error can be negligible when the false alarm rate is low. Another critical problem to calculate (\ref{eq:Jkz_Orig}) comes to the expectation operator taken with respect to state vector $\bf{x}_k$. This calculation can be performed off-line via Monte Carlo simulation by generating a set of target trajectory realizations and then calculating the averaged value. Another straightforward way is to use the mean value of the estimate of $\bf{x}_k$.
\subsection{Information gain factor arising from geometry-dependent target measurement uncertainty}
As shown in (\ref{eq:Jkz_xkmkGen}) , ${\bf{J}}_k^{\bf{Z}}({{\bf{x}}_k},{m_k})$ contains two parts: the first part is conventional information matrix  ${{\bf{H}}_k^T({{\bf{x}}_k}){\bf{R}}_k^{ - 1}({{\bf{x}}_k})} {{\bf{H}}_k}({{\bf{x}}_k})$ weighted by IRF ${\Lambda} _k^1({{\bf{x}}_k},{m_k})$ to account for MOU, its form is exactly the same as existing PCRLBs in  \cite{ref:8,ref:9,ref:15,ref:18}. The other part is the additional information matrix $\frac{{\partial \Psi ({{\bf{x}}_k})}}{{\partial {{\bf{x}}_k}}}{{\left[ {\frac{{\partial \Psi ({{\bf{x}}_k})}}{{\partial {{\bf{x}}_k}}}} \right]}^T}$ weighted by ${\Lambda} _k^2({{\bf{x}}_k},{m_k})$ , which quantifies the additional target Fisher information extracted from current measurements ${\bf{Z}}(k)$ due to the geometry-dependent TMU.  Considering the second part in (\ref{eq:Jkz_xkmkGen}) is essentially additional target Fisher information arising from geometry-dependent TMU, we define ${\Lambda} _k^2({{\bf{x}}_k},{m_k})$ as the information gain factor (IGF) to highlight the positive impact of considering dependence of TMU on T2R geometry when differentiating $\ln p({\bf{Z}}(k)|{{\bf{x}}_k},{m_k})$ w.r.t $\bf{x}_k$.

The conditional measurement information matrix derived in (\ref{eq:Jkz_xkmkGen}) is actually a generalized form that fully accounts for both MOU and geometry-dependent TMU. If we neglect the geometry-dependent TMU and treat both $P_d$ and $\bf{R}_k$ as state-independent parameters when differentiating $\ln p({\bf{Z}}(k)|{{\bf{x}}_k},{m_k})$ w.r.t $\bf{x}_k$, both $\bf{B}$ and $\bf{C}$ in (\ref{eq:CondLike_diff}) become zero matrices, hence ${\Lambda} _k^2({{\bf{x}}_k},{m_k})$ in (\ref{eq:Lambda_k2Orig}) equals to zero and the entire second part in (\ref{eq:Jkz_xkmkGen}) vanishes. Hence, ${\bf{J}}_k^{\bf{Z}}({{\bf{x}}_k},{m_k})$ in (\ref{eq:Jkz_xkmkGen}) reduces to the conventional form of existing PCRLBs in  \cite{ref:8,ref:9,ref:15,ref:18},
\begin{equation}\label{eq:Jkz_xkmkPCRLB}
{\bf{J}}_k^{\bf{Z}}({{\bf{x}}_k},{m_k}) = {\Lambda }_k^1({{\bf{x}}_k},{m_k}) {{\bf{H}}_k^T({{\bf{x}}_k}){\bf{R}}_k^{ - 1}({{\bf{x}}_k})} {{\bf{H}}_k}({{\bf{x}}_k})  
\end{equation}
However, if the TMU is considered to be partially dependent on the T2R geometry, i.e.,  $P_d$ is state-dependent but $\bf{R}_k$ is state-independent parameters when differentiating $\ln p({\bf{Z}}(k)|{{\bf{x}}_k},{m_k})$ w.r.t $\bf{x}_k$, $\bf{C}$ in (\ref{eq:CondLike_diff}) becomes zero matrix, the IGF $\Lambda _k^2({{\bf{x}}_k},{m_k})$ in (\ref{eq:Jkz_xkmkGen}) coincides to be the additional information impact parameter (AIIP) proposed in \cite{ref:17}, and equals to (\ref{eq:Lambda_AIIP}).
\begin{figure*}[htb]
\hrulefill
\vspace*{2pt}
\centering
\begin{equation}\label{eq:Lambda_AIIP}
\Lambda _k^2({{\bf{x}}_k},{m_k}) = {\left| {{{\bf{R}}_k}({{\bf{x}}_k})} \right|^{{m_k}/2}} \int\limits_{ - g}^g  \cdots  \int\limits_{ - g}^g  \cdots  \int\limits_{ - g}^g  \cdots  \int\limits_{ - g}^g {\frac{{ {{\Upsilon _1}}}}{{\beta ({{\bf{x}}_k},{m_k})}}{\rm{d}}{\bf{\hat z}}_k^1\left[ 1 \right] \cdots {\rm{d}}{\bf{\hat z}}_k^1\left[ n \right] \cdots } {\rm{d}}{\bf{\hat z}}_k^{{m_k}}\left[ 1 \right] \cdots {\rm{d}}{\bf{\hat z}}_k^{{m_k}}\left[ n \right]
\end{equation}
\hrulefill
\vspace*{-2pt}
\end{figure*}

\section{Application of IPCRLB to radar trajectory control}\label{sec:4}
The POMDP has been widely used as an online control framework for sensor management, in which the control decision-making problem is formulated as a first-order Markov process over time. Under POMDP, the best control command is selected via optimizing the objective function that quantitatively measures the goodness on the sensor-control resulted estimation. The IPCRLB derived in section \ref{sec:3} fundamentally formulates the impact of T2R geometry on the radar tracking accuracy and provides a much more accurate MSE bound for radar tracking in clutter with geometry-dependent TMU. To leverage the derived IPCRLB, we propose to formulate the trace of predictive IPCRLB as the objective function and minimize this cost function to select best trajectory-control command to optimize T2R geometry.  Compared to the state-of-the-art information-driven and task-driven methods, the proposed approach avoids operating complicated pseudo tracking using enumerated PIMS, and delivers intensively improved tracking results.

Since the TMU in bistatic radar is more sensitive to the T2R geometry than that in monostatic radar, to better demonstrate trajectory optimization benefits, we concentrate on bistatic radar system with stationary transmitter and control the receiver trajectory to optimize the transmitter-target-receiver geometry. Recall the bistatic radar system depicted in 2D Cartesian coordinate in section \ref{sec:2}, the full state of receiver consists of 2D position and velocity component, denoted by ${{\bf{x}}_{R,k}} = {\left[ {{x_{R,k}}{\kern 4pt} {y_{R,k}}{\kern 4pt} {{\dot x}_{R,k}}{\kern 5pt} {{\dot y}_{R,k}}} \right]^T}{\rm{ = }}{\left[ {{{\left( {{\bf{x}}_{R,k}^p} \right)}^T}{\kern 2pt} {{\left( {{\bf{x}}_{R,k}^v} \right)}^T}} \right]^T}$,  with the receiver trajectory represented by its position component ${{\bf{x}}_{R,k}^p}$ over time. In realistic scenario, to guarantee tracking performance,  the radar usually operates in a very fast sampling rate, we follow \cite{ref:21}-\cite{ref:24} and assume the receiver moves with a nearly constant velocity (NCV) model during the interval of two consecutive scans, denoted by ${T_k}={t_k}-{t_{k-1}}$.  Therefore, the receiver velocity vector at any time ${t_l} \in ({t_{k - 1}},{\kern 2pt}{t_k}]$ keeps same and equals to the velocity vector at last time $t_{k-1}$, that is
\begin{equation}\label{eq:Xv_NCV}
{\bf{x}}_{R,l}^v = {\bf{x}}_{R,k - 1}^v = \left[ \begin{array}{l}
{{\dot x}_{R,k - 1}}\\
{{\dot y}_{R,k - 1}}
\end{array} \right] = \left[ \begin{array}{l}
v_{k - 1}^R\cos (w_{k - 1}^R)\\
v_{k - 1}^R\sin (w_{k - 1}^R)
\end{array} \right]
\end{equation}
where $v_{k - 1}^R$ and $w_{k - 1}^R$ denote the velocity and heading angle of receiver velocity vector at time $t_{k-1}$, respectively. Hence, the receiver position at time $t_k$ is collectively determined by its position and velocity at last time $t_{k-1}$, i.e.,
\begin{equation}\label{eq:Xp_NCV}
{\bf{x}}_{R,k}^p = \left[ \begin{array}{l}
{x_{R,k}}\\
{y_{R,k}}
\end{array} \right] = \left[ \begin{array}{l}
{x_{R,k - 1}} + v_{k - 1}^RT_k\cos (w_{k - 1}^R)\\
{y_{R,k - 1}} + v_{k - 1}^RT_k\sin (w_{k - 1}^R)
\end{array} \right]
\end{equation}
As shown above, the receiver velocity $v_{k - 1}^R$ and heading angle $w_{k - 1}^R$ at last time $t_{k-1}$ determine the receiver trajectory position ${{\bf{x}}_{R,k}^p}$ at current time $t_k$. Let's define the receiver trajectory control command at $t_{k-1}$ as ${{\bf{u}}_{k - 1}} = {\left[ {v_{k - 1}^R{\kern 1pt} {\kern 1pt} {\kern 1pt} {\kern 1pt} {\kern 1pt} {\kern 1pt} w_{k - 1}^R} \right]^T}$. To highlight the dependence of receiver state at time $t_k$ on the control command at $t_{k-1}$, we define ${{\bf{x}}_{R,k}} \buildrel \Delta \over = {{\bf{x}}_{R,k}}\left( {{{\bf{u}}_{k - 1}}} \right)$. In consideration of realistic maneuver capability of radar platform, both the value and changing rate of receiver velocity and its heading angle need to satisfy the following constraints 
\begin{equation}\label{eq:Command_Constrain}
\left\{ \begin{array}{l}
{\kern 20pt} v_{min}^R \le v_{k - 1}^R \le v_{max }^R,{\kern 5pt} w_{k - 1}^R \le w_{max }^R\,\\
\begin{array}{*{20}{c}}
{v_{k - 2}^R - a_{max }^vT_k \le v_{k - 1}^R \le v_{k - 2}^R - a_{max }^vT_k}\\
{w_{k - 2}^R - a_{max }^wT_k \le w_{k - 1}^R \le w_{k - 2}^R + a_{max }^wT_k}
\end{array}
\end{array} \right.
\end{equation}
Note that the receiver platform is usually airborne and needs to keep minimum velocity $v_{min}^R$ to avoid crash, $v_{max }^R$ and $w_{max }^R$ are the maximum velocity and heading angle of receiver platform,  $a_{max }^v$ and $a_{max }^w$ are the maximum acceleration and angular rate of receiver platform. ${{\bf{u}}_{k - 2}} = {\left[ {v_{k - 2}^R{\kern 6pt} w_{k - 2}^R} \right]^T}$denotes the receiver trajectory control command at last time $t_{k-2}$.  (\ref{eq:Command_Constrain}) indicates ${{\bf{u}}_{k - 1}}$ takes value in a continuous space. For sake of simplicity, we discretize the value space of ${{\bf{u}}_{k - 1}}$ into a finite set of admissible commands, i.e., 
${{\bf{U}}_{k - 1}}{\rm{ = }}\left\{ {{\bf{u}}_{k - 1}^i} \right\}_{i = 1}^{({N_v+1})({N_w+1})}$, with ${\bf{u}}_{k - 1}^i{\rm{ = }}{\left[ {v_{k - 1,i}^R{\kern 1pt} {\kern 1pt} {\kern 1pt} {\kern 1pt} {\kern 1pt} {\kern 1pt} w_{k - 1,i}^R} \right]^T}$ denoting the $i$-th command. Each element in ${\bf{u}}_{k - 1}^i$ needs to satisfy the constraints in (\ref{eq:Command_Constrain}) and can be calculated by
\begin{equation}\label{eq:Command_Library}
\left\{ {\begin{array}{*{20}{c}}
{v_{k - 1,i}^R{\rm{ = }}v_{k - 2}^R - a_{max }^v{T_k}{\rm{ + }}{\varsigma _i}\frac{{2a_{max }^v{T_k}}}{{N_v}}{\kern 3pt}{\varsigma _i} = 0,...,{N_v}{\rm{        }}}\\
{w_{k - 1,i}^R = w_{k - 2}^R - a_{max }^w{T_k} + {\zeta _j}\frac{{2a_{max }^w{T_k}}}{{{N_w}}}{\rm{,}}{\kern 6pt}{\zeta _j} = 0,...,{N_w}{\rm{        }}}
\end{array}} \right.
\end{equation}
where $N_v$ and $N_w$ are the discretized numbers of receiver velocity and heading angle respectively. 

We consider an explicit decision-making problem that selects the best control command ${{\bf{u}}_{k - 1}}$ at time $t_{k-1}$ to obtain proper receiver state ${{\bf{x}}_{R,k}}\left( {{{\bf{u}}_{k - 1}}} \right)$, such that the transmitter-target-receiver geometry at time $t_k$ can be optimized.  Based on the designated ${{\bf{x}}_{R,k}}\left( {{{\bf{u}}_{k - 1}}} \right)$,  the receiver is actuated accordingly to acquire high-quality target measurement, thereby improving tracking accuracy.  The myopic control policy is considered at the rest of this section, the term “myopic” means only deciding one control command one-step ahead rather than planning multiple commands multi-step into the future.  To cast the receiver trajectory control problem into the POMDP framework, several other elements need to be defined as follows:
\begin{itemize}[label=\textbullet, font=\Large]
\item $p({\bf{x}}_k|{\bf{Z}}^{k-1})$: predicted density of target state $\bf{x}_k$ given measurements up to time $t_{k-1}$
\item ${\bf{U}}_{k-1}$: a set of finite admissible control commands for radar receiver to operate at time $t_{k-1}$
\item  $p({\bf{z}}_k|{\bf x}_k,{{\bf{x}}_{R,k}}\left( {{{\bf{u}}_{k - 1}}} \right))$: the likelihood of measurement ${\bf{z}}_k$ given target state $\bf{x}_k$ and command ${\bf{u}}_{k - 1}$ resulted receiver state ${{\bf{x}}_{R,k}}\left( {{{\bf{u}}_{k - 1}}} \right)$ at time $t_k$
\item $C\left( {{{\bf{x}}_k},{{\bf{x}}_{R,k}}\left( {{{\bf{u}}_{k - 1}}} \right)} \right)$: a real-valued objective function associated with command ${\bf{u}}_{k - 1}$ resulted receiver state ${{\bf{x}}_{R,k}}\left( {{{\bf{u}}_{k - 1}}} \right)$ and target state $\bf{x}_k$
\end{itemize}

For the objective function, we utilize the trace of predictive IPCRLB as the cost function to measure the MSE error of trajectory-control resulted target state estimate and find the best control command that minimizes this cost function. At time $t_{k}$, the target state $\bf{x}_k$ is  characterized by its predicted distribution $p({\bf{x}}_k|{\bf{Z}}^{k-1})$, thus the best command ${{{\bf{u}}_{k - 1}}}$ is commonly chosen by minimizing the statistical mean of the cost function $C\left( {{{\bf{x}}_k},{{\bf{x}}_{R,k}}\left( {{{\bf{u}}_{k - 1}}} \right)} \right)$ over all realizations of $\bf{x}_k$, hence, we can formulate the receiver trajectory control problem as 
\begin{equation}
\begin{split}
 {{\bf{u}}_{k - 1}} & = \mathop {\rm min }\limits_{{\bf{u}}_{k - 1}^i \in {{\bf{U}}_{k - 1}}} {E_{{{\bf{x}}_k}}}\left\{ {C\left( {{{\bf{x}}_k},{{\bf{x}}_{R,k}}\left( {{{\bf{u}}_{k - 1}^i}} \right)} \right)} \right\}\\
&  = \mathop {\rm min }\limits_{{\bf{u}}_{k - 1}^i \in {{\bf{U}}_{k - 1}}} {{Tr}}{\left( {{\bf{J}}_k^{ - 1}} \right)_{{{\bf{x}}_k} = {{{\bf{\hat x}}}_{k|k - 1}}}}
\end{split}
\end{equation}
with elements of each command ${\bf{u}}_{k - 1}^i$ constrained by (\ref{eq:Command_Constrain}). 
\begin{figure}[htbp]
\centering
\includegraphics[width=0.48\textwidth]{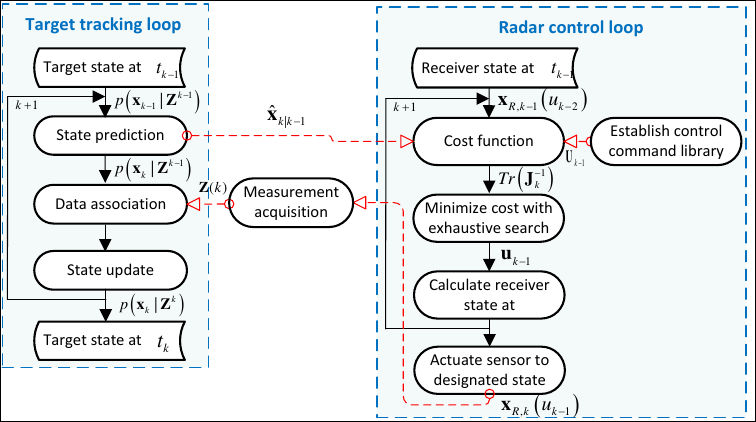}

\caption{Receiver trajectory control framework}
\label{fig:RadarTraOpt}
\end{figure}

The receiver trajectory control framework is shown in Fig. \ref{fig:RadarTraOpt}, in which the interaction between trajectory control loop and target tracking loop is clearly demonstrated. Based upon the outputs at time $t_{k-1}$ from tracking loop (target state estimate $p({\bf{x}}_{k-1}|{\bf{Z}}^{k-1})$ ) and control loop (receiver state ${{\bf{x}}_{R,k-1}}\left( {{{\bf{u}}_{k - 2}}} \right)$), main steps of receiver trajectory control are summarized as follows:
\begin{itemize}[label=\textbullet, font=\Large]
\item step 1: propagate $p({\bf{x}}_{k-1}|{\bf{Z}}^{k-1})$ through target dynamic model to obtain predicted state estimate $p({\bf{x}}_{k}|{\bf{Z}}^{k-1})$ at time $t_{k}$. Note that the target tracking considered in this paper assumes target track has been successfully initiated and concentrates on the track maintenance by implementing data association and filtering update to obtain target posterior state estimate at each time.
\item step 2: based upon receiver state ${{\bf{x}}_{R,k-1}}\left( {{{\bf{u}}_{k - 2}}} \right)$ , generate each admissible command via (\ref{eq:Command_Library}) and establish the discretized receiver trajectory control command library ${{\bf{U}}_{k - 1}}{\rm{ = }}\left\{ {{\bf{u}}_{k - 1}^i} \right\}_{i = 1}^{({N_v+1})({N_w+1})}$, with each element of ${{\bf{u}}_{k - 1}^i}$ satisfied the platform maneuver constraints shown in (\ref{eq:Command_Constrain}), .
\item step 3: calculate the cost function ${{Tr}}{\left( {{\bf{J}}_k^{ - 1}} \right)_{{{\bf{x}}_k} = {{{\bf{\hat x}}}_{k|k - 1}}}}$ that evaluates at the mean of predicted state estimate $p({\bf{x}}_{k}|{\bf{Z}}^{k-1})$. 
\item step 4: select the best control command ${{\bf{u}}_{k - 1}}$ that minimizes ${{Tr}}{\left( {{\bf{J}}_k^{ - 1}} \right)_{{{\bf{x}}_k} = {{{\bf{\hat x}}}_{k|k - 1}}}}$ using exhaustive search inside the discrete control command space ${{\bf{U}}_{k - 1}}$ .
\item step 5: based on the selected command ${{\bf{u}}_{k - 1}}$, calculate the full state of receiver ${{\bf{x}}_{R,k}}\left( {{{\bf{u}}_{k - 1}}} \right)$ by (\ref{eq:Xv_NCV}) and (\ref{eq:Xp_NCV}).
\item  step 6: actuate receiver to the designated state ${{\bf{x}}_{R,k}}\left( {{{\bf{u}}_{k - 1}}} \right)$ to acquire high quality radar measurements ${\bf Z}(k)$ at time $t_k$, then fed ${\bf Z}(k)$ back to the target tracking loop to implement data association and state update to calculate the posterior estimate $p({\bf{x}}_{k}|{\bf{Z}}^{k})$ at time $t_k$.
\end{itemize}

\section{Simulation and Discussion}\label{sec:5}
In this section, we consider a scenario deploying a bistatic radar to track a moving target in clutter in 2D surveillance space. The advanced television systems committee (ATSC) signal is selected as the illuminated signal, the FIM for time delay and Doppler shift of ATSC signal is then given by \cite{ref:34}
\begin{equation}\label{eq:Rk_Simp}
\begin{split}
&{{\bf{J}}_{ATSC}}({\tau _k},{\xi _k}) = \Psi \left( {{\Delta _k}} \right) \\ 
& \times \left[ {\begin{array}{*{20}{c}}
{\frac{{2{\alpha ^2}}}{{T_{Sym}^2}}(\frac{{ - 1}}{{{\pi ^2}}} + \frac{1}{{96{\alpha ^2}}} + \frac{1}{8})}&0\\
0&{2T_{Sym}^2(\frac{1}{{4\alpha }} + \frac{{{N^2} - 1}}{3})}
\end{array}} \right]
\end{split}
\end{equation}
where ${T_{Sym}}$ is the symbol period of the ATSC signal, $\alpha $ is the roll-off factor, and $N$ is the number of transmitted symbols. By substituting the diagonal elements of matrix at the right hand side of (\ref{eq:Rk_Simp}) to (\ref{eq:Rk_Final}), the full expressions of ${{\bf{R}}_k}\left( {{\Delta _k}} \right)$ can be obtained. Following \cite{ref:34},  we set $\alpha  = 0{.}05762$, ${T_{Sym}} = 93ns$, $N = 1076000$, ${f_c} = 63{.}1 MHz$, ${\sigma _{{\theta _0}}} = {3^o}$.  Reminder of this section conducts three case studies to validate the effectiveness and superiority of model and methods proposed at section~\ref{sec:2}, \ref{sec:3} and \ref{sec:4}, respectively.
\subsection{Case 1: impact of T2R geometry on target measurement uncertainty}
This case study aims to investigate the impact of T2R geometry ${\Delta_k}$ on $P_d$ and ${\bf{R}}_k$ via (\ref{eq:Pd_xk}) and (\ref{eq:Rk_Final}) proposed in section~\ref{sec:2}. Since the location and shape of triangle ${\Delta_k}$ are uniquely described by $\left( {{d_k},{L_k},{\theta _k},{\theta _{TR,k}}} \right)$, to obtain diverse T2R geometries, we set both the transmitter and receiver to be static with ${L_k} = 5km$ and  ${\theta _{RT,k}} = \pi$, and control target moving to generate different target range ${R_{R,k}}$ or DOA ${\theta _k}$.  The target velocity is set to be a constant $50m/s$, bistatic angle ${\beta _k}$ equals to $45^0$,  the false alarm rate and transmitted signal constant are set to be ${P_{FA}} = 0{.}001$ and  ${{\vartheta _0}} = 5000$, respectively.

\begin{figure*}[htbp]
\centering
\subfloat[${\sigma_d}$ versus ${\theta _k}$]{\includegraphics[width=0.25\textwidth]{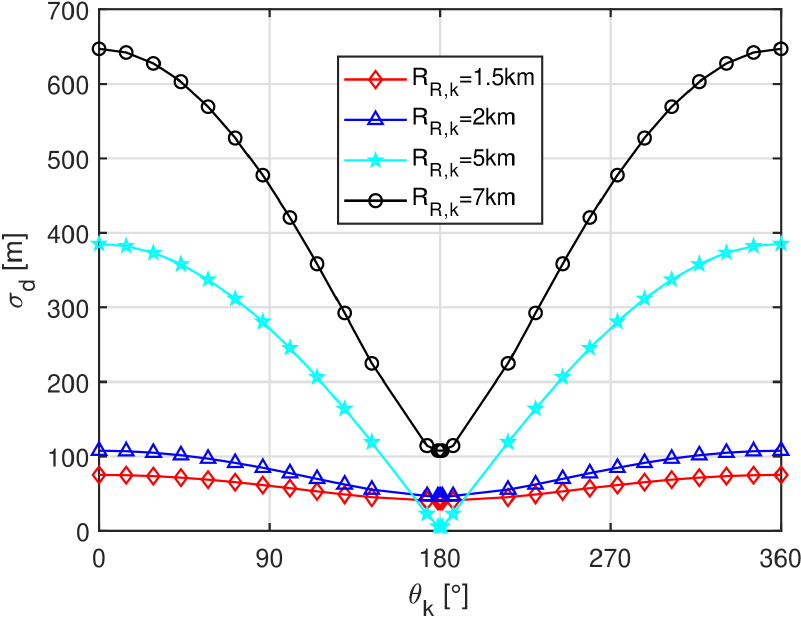}%
\label{fig:Sigmad_theta-2}}
\subfloat[${\sigma_v}$ versus ${\theta _k}$]{\includegraphics[width=0.25\textwidth]{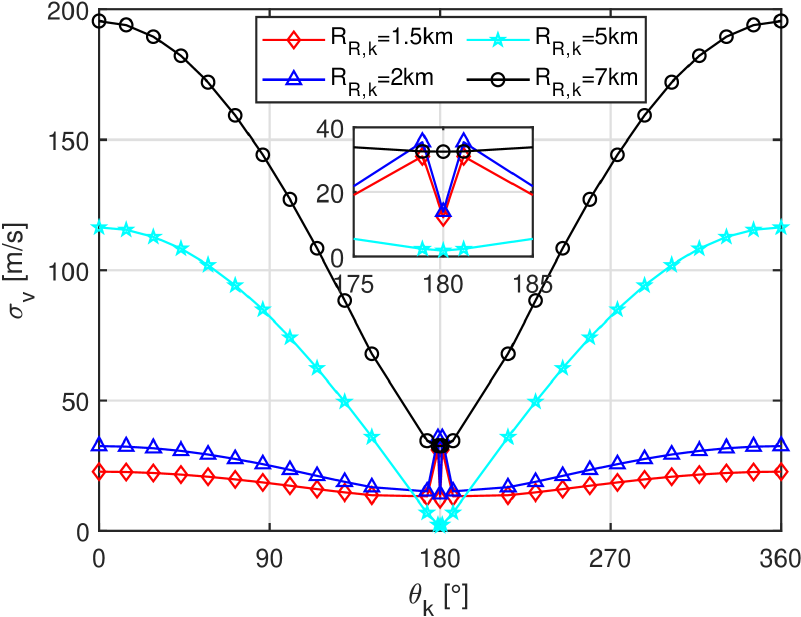}%
\label{fig:SigmaV_theta-2}}
\subfloat[${\sigma_\theta}$ versus ${\theta _k}$]{\includegraphics[width=0.25\textwidth]{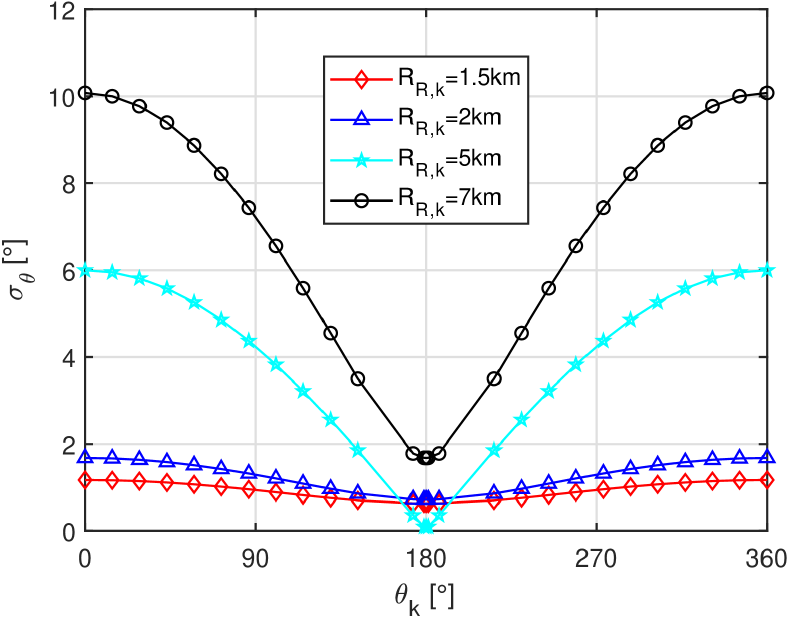}%
\label{fig:SigmaTheta_theta-2}}
\subfloat[${P_d}$ versus ${\theta _k}$]{\includegraphics[width=0.25\textwidth]{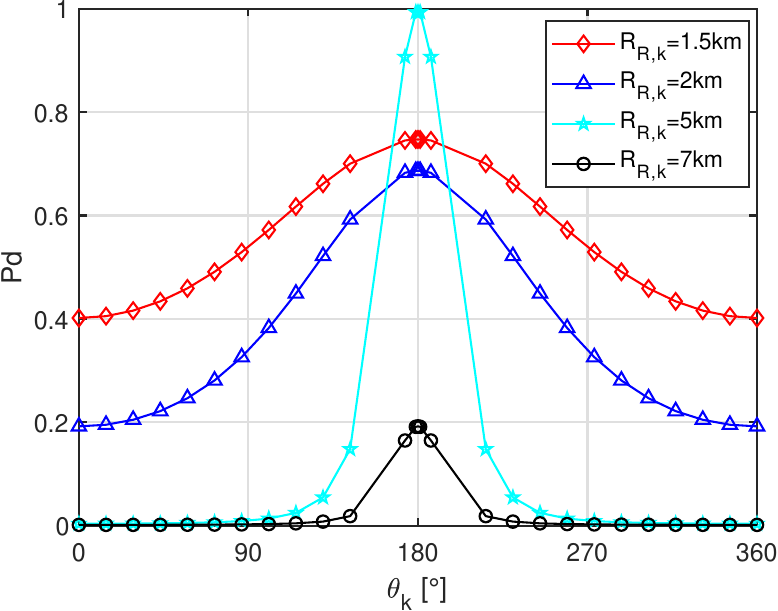}%
\label{fig:Pd_theta-2}}
\caption{TMU varies as target DOA ${\theta_k}$ w.r.t. different target range ${R_{R,k}}$.}
\label{fig:TMU_theta-2}
\end{figure*}

\begin{figure*}[htbp]
\centering
\subfloat[${\sigma_d}$ versus ${R_{R,k}}$]{\includegraphics[width=0.25\textwidth]{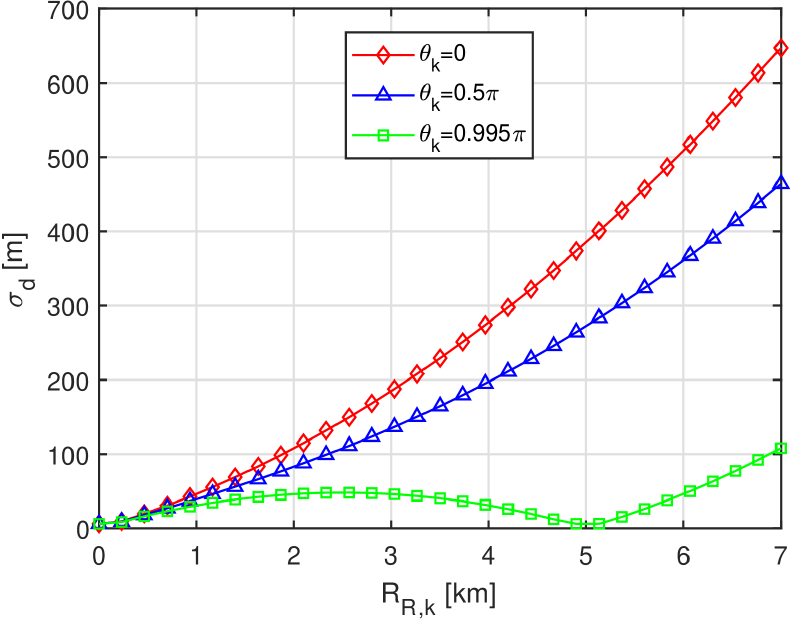}%
\label{fig:Sigmad_Rk-2}}
\subfloat[${\sigma_v}$ versus ${R_{R,k}}$]{\includegraphics[width=0.25\textwidth]{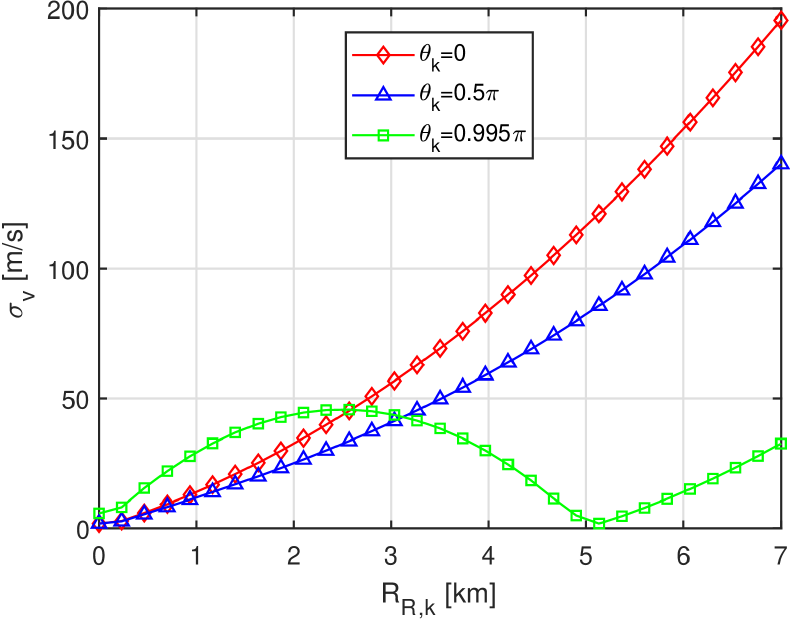}%
\label{fig:SigmaV_Rk-2}}
\subfloat[${\sigma_\theta}$ versus ${R_{R,k}}$]{\includegraphics[width=0.25\textwidth]{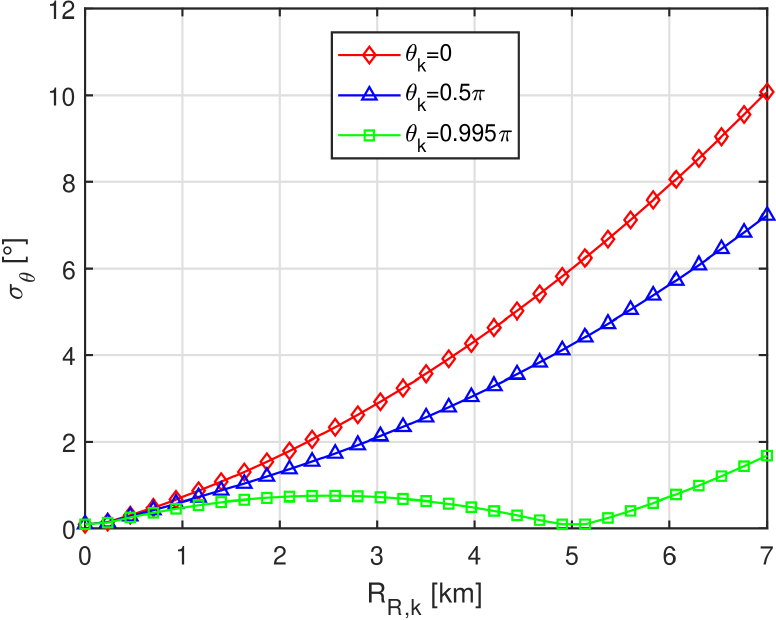}%
\label{fig:SigmaTheta_Rk-2}}
\subfloat[${P_d}$ versus ${R_{R,k}}$]{\includegraphics[width=0.25\textwidth]{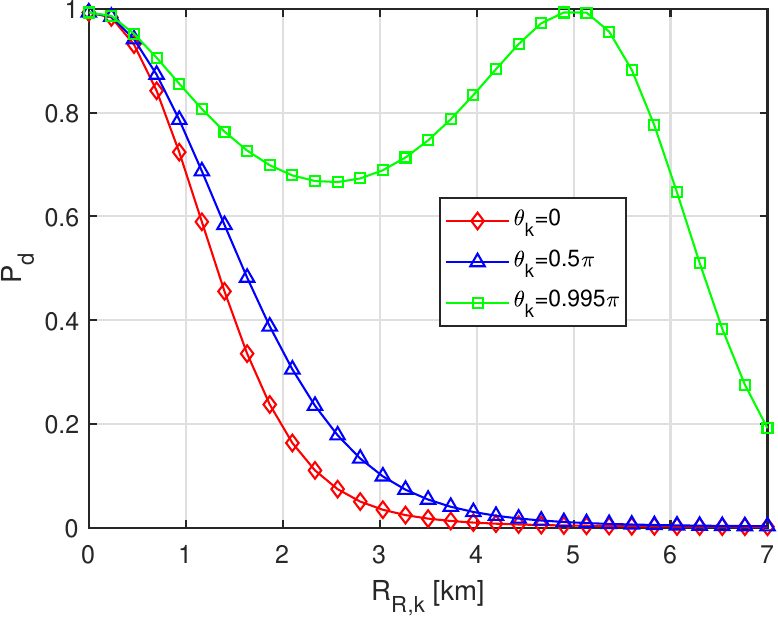}%
\label{fig:Pd_Rk-2}}
\caption{TMU varies as target range ${R_{R,k}}$ w.r.t. different target DOA ${\theta_k}$.}
\label{fig:TMU_Rk-2}
\end{figure*}

As shown in Fig. \ref{fig:TMU_theta-2}, the TMU in terms of measurement error standard deviation ${\sigma _d}$, ${\sigma _v}$, ${\sigma _\theta }$ and detection probability ${P_d}$ varies evidently as the DOA ${\theta _k}$ changes from ${0^o}$ to ${360^o}$ w.r.t. different target range ${R_{R,k}}$. At ${\theta _k}={180^o}$, ${\sigma _d}$, ${\sigma _v}$ and ${\sigma _\theta }$ reach their minimums while ${P_d}$ reaches its maximum, both indicating the least TMU. This is because when ${\theta _k}={180^o}$, the transmitter, target and receiver are collinear, the value of ${R_{T,k}}{R_{R,k}}$ in (\ref{eq:SNR}) decreases to the minimum such that the echo SNR ${\Psi}$ is maximized. Notably, ${\sigma _v}$ slightly increases near ${\theta _k}={180^o}$ due to the resolution loss in Doppler blind zone. Fig. \ref{fig:TMU_Rk-2} shows that the TMU changes dramatically with varied target range ${R_{R,k}}$ w.r.t. different target DOA ${\theta _k}$. It can be observed there that target measurement achieves the least uncertainty when target moves towards the transmitter-receiver baseline. Therefore, these numerical results validate the TMU in terms of both measurement error covariance and detection probability is indeed highly dependent on the T2R geometry. 

\subsection{Case 2: comparison of IPCRLB with state-of-the-art bounds}
To validate the superiority of the proposed IPCRLB for radar tracking in clutter with geometry-dependent TMU, two state-of-the-art bounds are implemented and compared. They are the conventional PCRLB \cite{ref:8,ref:9,ref:15,ref:18} and the EFIM \cite{ref:17}, respectively.  To investigate the impact of T2R geometry on the compared bounds, following the same setting as case 1, the target is controlled to move along its trajectory with varying DOA ${\theta _k}$ or range ${R_{R,k}}$ to generate different T2R geometries. 

\begin{figure*}[htbp]
\centering
\subfloat[${R_{R,k}}= $1.5$ km$]{\includegraphics[width=0.25\textwidth]{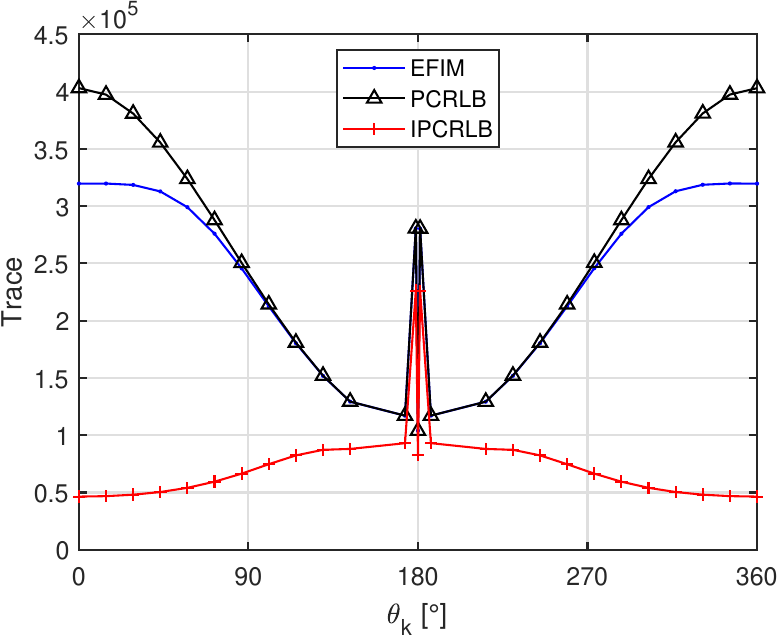}%
\label{fig:PCRLB-Theta-1.5km}}
\subfloat[${R_{R,k}}=2km$]{\includegraphics[width=0.25\textwidth]{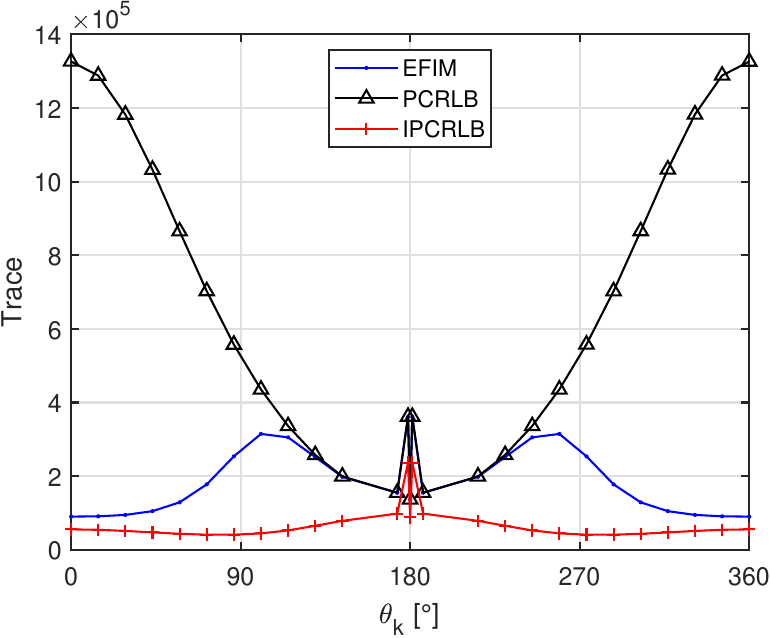}%
\label{fig:PCRLB-Theta-2km}}
\subfloat[${R_{R,k}}=5km$]{\includegraphics[width=0.25\textwidth]{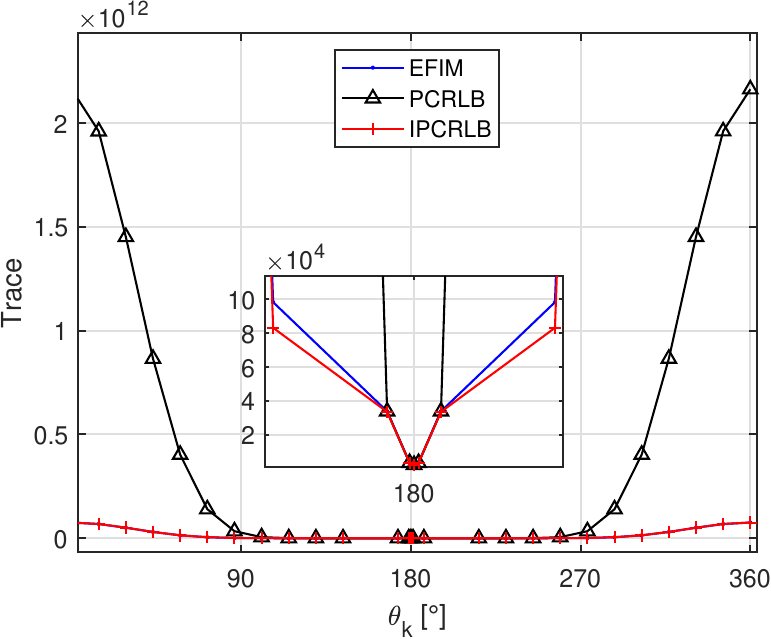}%
\label{fig:PCRLB-Theta-5km}}
\subfloat[${R_{R,k}}=7km$]{\includegraphics[width=0.25\textwidth]{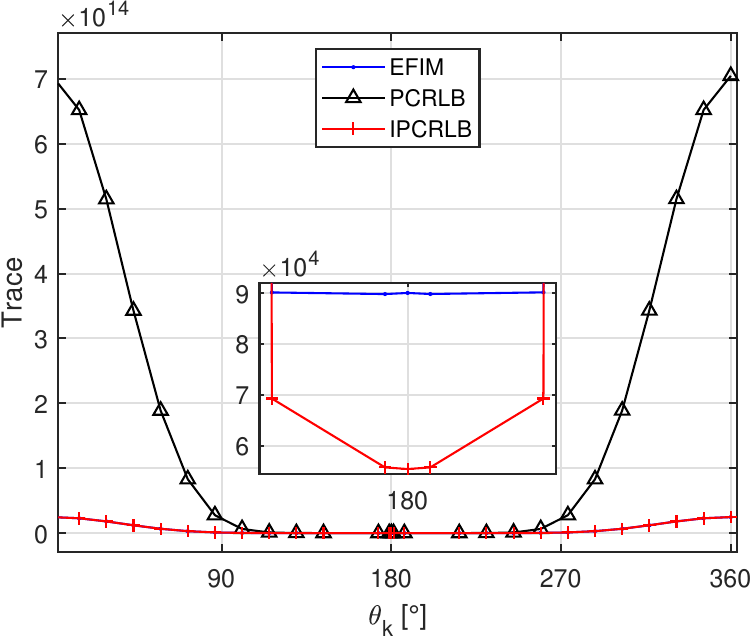}%
\label{fig:PCRLB-Theta-7km}}
\caption{Comparison of the trace of different bounds versus varied target DOA ${\theta_k}$.}
\label{fig:PCRLB_theta}
\end{figure*}

\begin{figure*}[htbp]
\centering
\subfloat[$\theta_k=0$]{\includegraphics[width=0.25\textwidth]{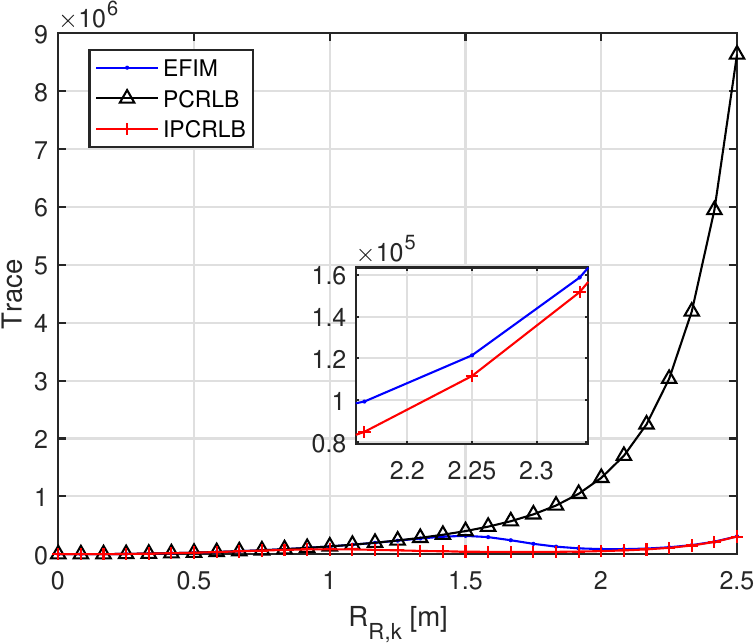}%
\label{fig:PCRLB-Rk-0pi}}
\subfloat[$\theta_k= $0.5$ \pi$]{\includegraphics[width=0.25\textwidth]{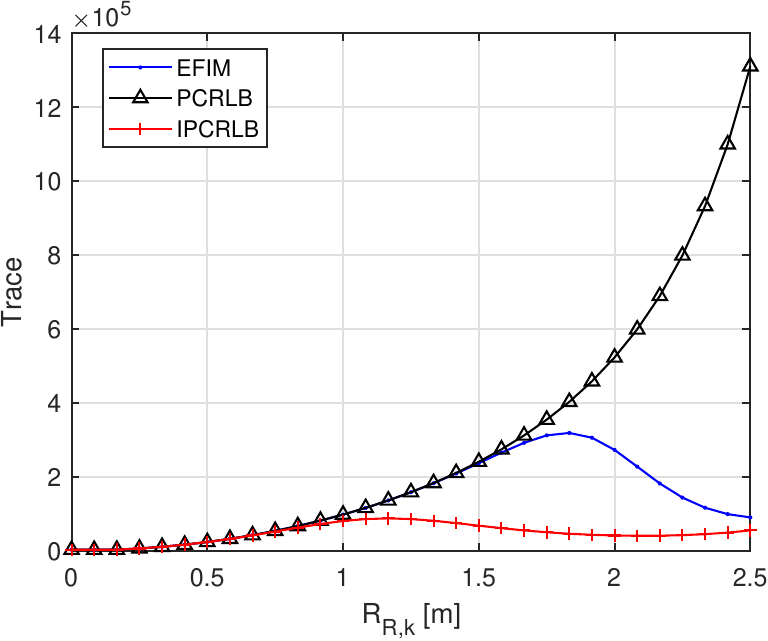}%
\label{fig:PCRLB-Rk-0.5pi}}
\subfloat[$\theta_k= $0.99$ \pi$]{\includegraphics[width=0.25\textwidth]{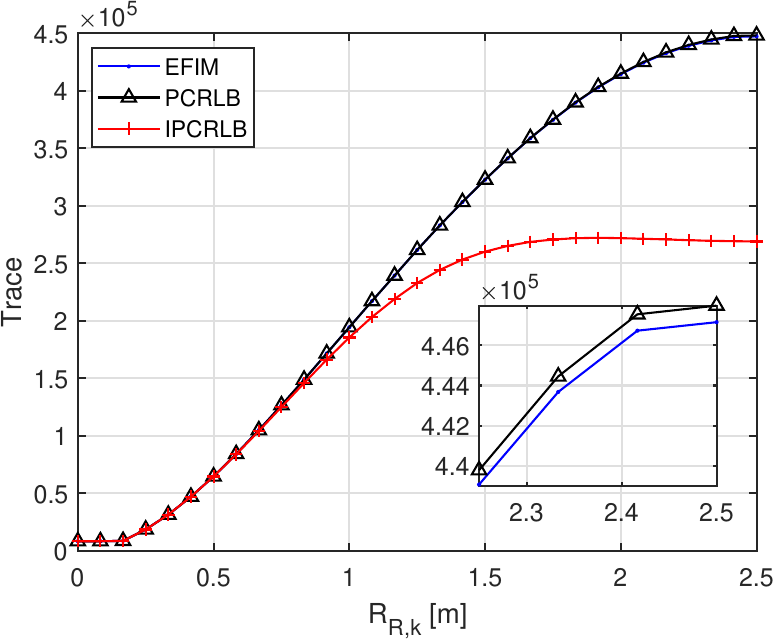}%
\label{fig:PCRLB-Rk-0.99pi}}
\caption{Comparison of trace of different bounds versus varied target range $R_{R,k}$.}
\label{fig:PCRLB_Rk}
\end{figure*}

\begin{figure}[htbp]
\centering 
\subfloat[Trace of compared PCRLBs versus $P_{FA}$]{\includegraphics[width=0.35\textwidth]{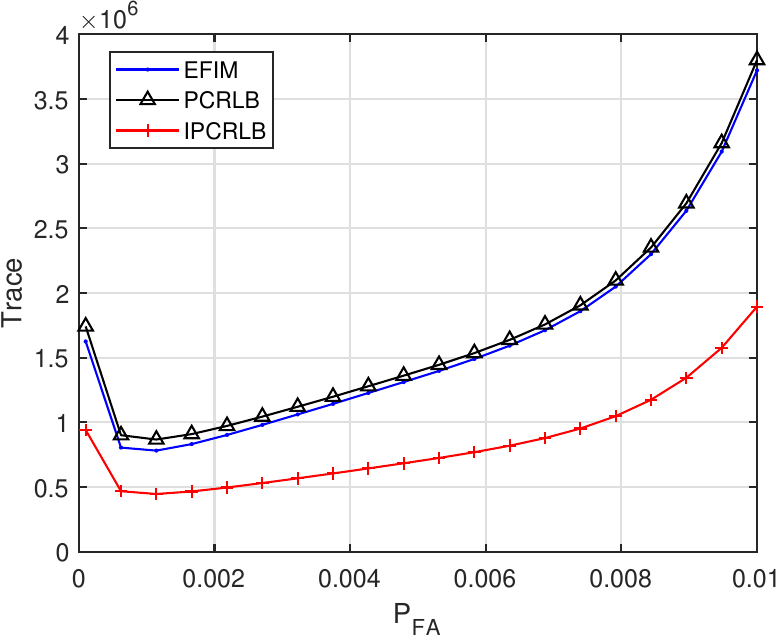}\label{fig:PCRLB-Pfa}}
\hfill
\subfloat[Trace of compared PCRLBs versus ${\vartheta_0}$]{\includegraphics[width=0.35\textwidth]{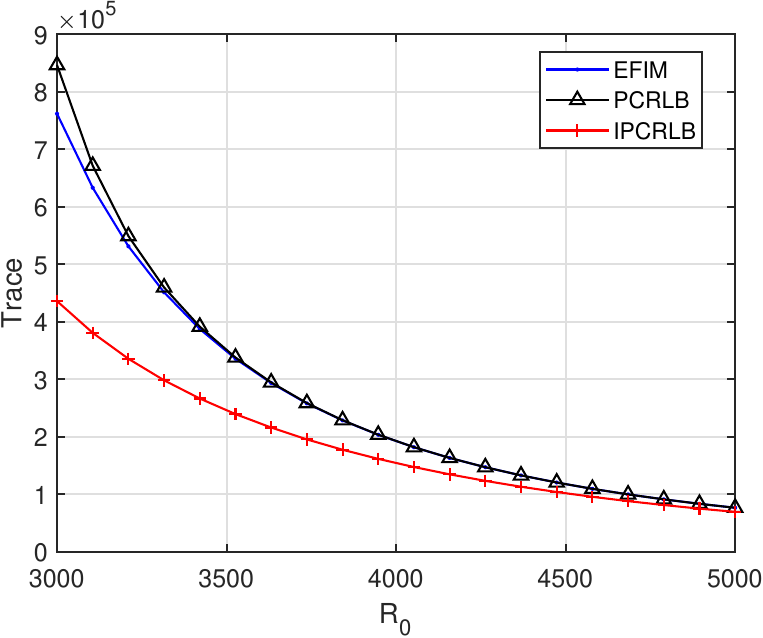}\label{fig:PCRLB-R0}}
\caption{Comparison of trace of different bounds versus varied false alarm rate and transmitted signal constant}
\label{fig:PCRLB_Comp}
\end{figure}

Comparisons of the trace of three bounds versus varied ${\theta _k}$ and ${R_{R,k}}$ are demonstrated in Fig.~\ref{fig:PCRLB_theta} and \ref{fig:PCRLB_Rk} respectively. As shown there, the proposed IPCRLB delivers the lowest MSE bound, followed by the EFIM, with conventional PCRLB the highest bound. This is because IPCRLB extracts more additional target Fisher information by fully considering both $P_d$ and ${\bf{R}}_k$ as state-dependent parameters when differentiating $\ln p({\bf{Z}}(k)|{{\bf{x}}_k},{m_k})$ w.r.t $\bf{x}_k$. In contrast, EFIM ignores the dependence of ${\bf{R}}_k$ on T2R geometry and PCRLB completely treats TMU as state-independent parameters, leading to increased (over-conservative) MSE bound. Furthermore, as shown in Fig. \ref{fig:PCRLB-Theta-1.5km}, Fig. \ref{fig:PCRLB-Theta-2km} and Fig.\ref{fig:PCRLB_Rk}, the improvement of IPCRLB over PCRLB and EFIM increases as TMU rises (i.e., ${\sigma _d}$, ${\sigma _v}$ and ${\sigma _\theta }$ increase while $P_d$ decreases as shown in Fig. \ref{fig:TMU_theta-2} and Fig. \ref{fig:TMU_Rk-2}). However, if TMU is too high, the improvement of IPCRLB over EFIM vanishes as indicated in  Fig. \ref{fig:PCRLB-Theta-5km}  and  Fig. \ref{fig:PCRLB-Theta-7km}. Moreover, similarly improved results of proposed IPCRLB over compared bound w.r.t. varied false alarm rate and transmitted signal constant can be observed in Fig.\ref{fig:PCRLB-Pfa} and Fig.\ref{fig:PCRLB-R0}, respectively. In consequence, numerical results of this case study verify the superiority of proposed IPCRLB compared to state-of-the-art bounds.

\subsection{Case 3: superiority of IPCRLB applied to receiver trajectory control}
To validate the superiority of proposed IPCRLB formulating MSE lower bound for radar tracking in clutter with geometry-dependent TMU, the trace of predictive IPCRLB is utilized as the cost function and minimized to optimize the T2R geometry to acquire high-quality target measurement for improved radar tracking in clutter. For simplicity, this control method is abbreviated as min tr(IPCRLB). Several state-of-the-art radar trajectory control methods are also implemented and compared:
\begin{itemize}
    \item max KLD: a representative information-driven method, the Kullback-Leibler divergence(KLD) is formulated as a reward function and maximized to select the best receiver moving control command \cite{ref:21}. 
    \item min PDST: a representative task-driven method, the posterior distance between sensor and target(PDST) is formulated as a cost function and minimized to select the best receiver moving control command \cite{ref:32}.
    \item min tr(PCRLB): bound based method, the trace of  predictive PCRLB is utilized as the cost function and minimized to select the best receiver moving control command. 
    \item     fixed: the receiver keeps fixed at its original position. 
    \item     random: randomly choose the command to control receiver moving trajectory. 
\end{itemize}

Transmitter is fixed at the origin, initial state of receiver equals ${{\bf{x}}_{R,0}} = {\left[ {5000m {\kern 4pt} 0m {\kern 4pt} 0m/s {\kern 4pt} 0m/s} \right]^T}$, the target moves with a nearly constant velocity model with initial state ${{\bf{x}}_{0}} = {\left[ {5000m {\kern 4pt} 5000m {\kern 4pt} -80m/s {\kern 4pt} -100m/s} \right]^T}$, the minimum and maximum velocity of receiver platform are set be  $v_{min}^R = 1m/s$ and $v_{max }^R=100m/s$, the maximum acceleration and angular rate of receiver platform are set to be $a_{max }^v=5m/s^2$ and $a_{max }^w=30^0/s^2$, discretized numbers of receiver platform velocity and heading angle are set to be $N_v=40$ and $N_w=20$. ${{\vartheta _0}}$ is set to be 9000, the experiment repeats for 200 Monte Carlo runs. Other parameters are set the same as case 1. 

\begin{figure}[htbp]
\centering
{\includegraphics[width=0.35\textwidth]{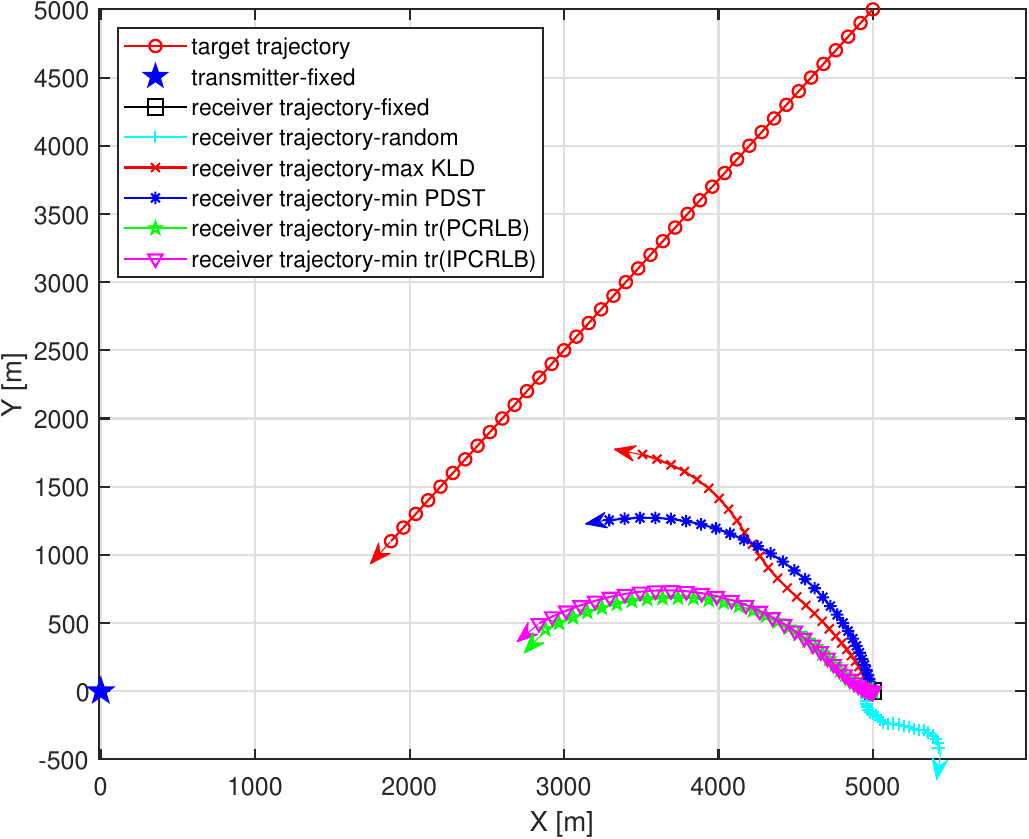}%
\caption{Comparison of T2R Geometries of different receiver trajectory control methods.}
\label{fig:OptReceiverTrajectory}}
\end{figure}

\begin{figure*}[htbp]
\centering
{\includegraphics[width=1\textwidth]{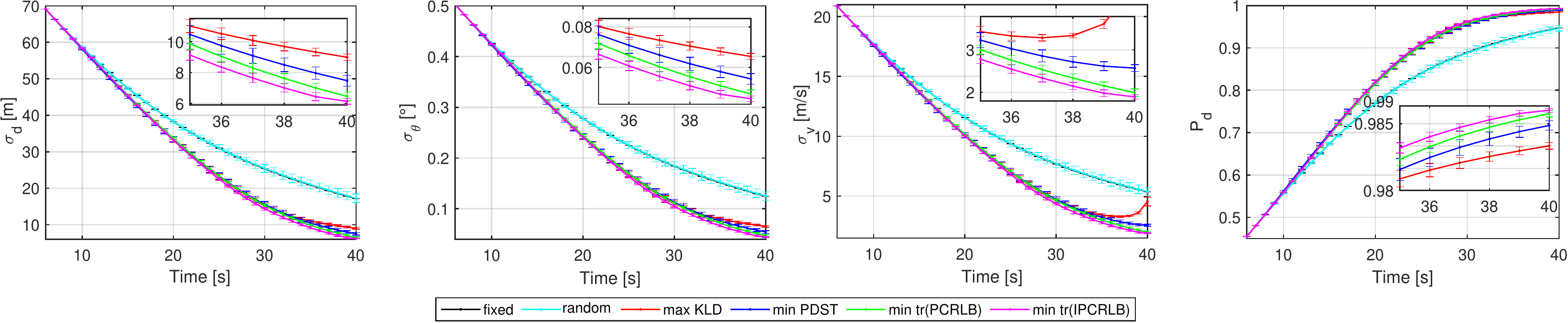}%
}
\caption{Comparison of target measurement uncertainties among different receiver trajectory control methods.}
\label{fig:RkSNR_Opt}
\end{figure*}

\begin{figure}[htbp]
\centering
{\includegraphics[width=0.35\textwidth]{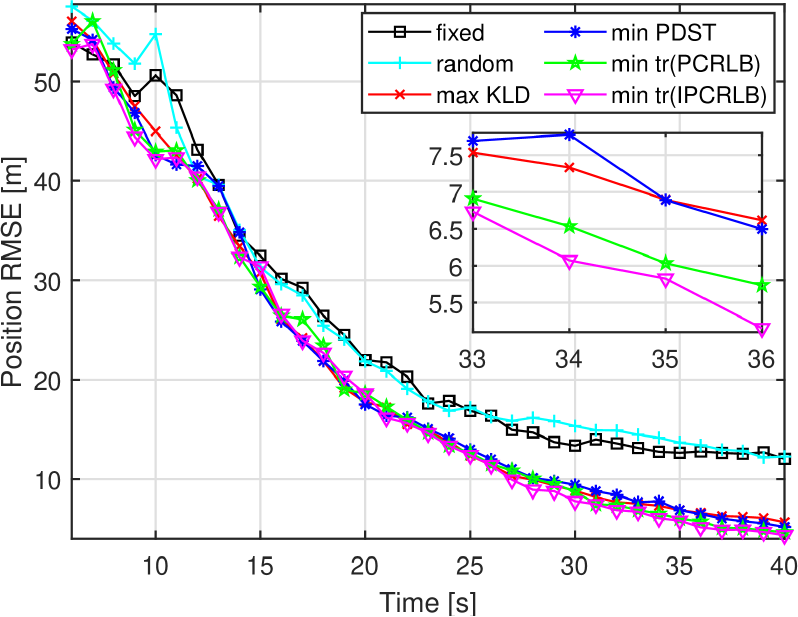}%
\caption{Comparison of position RMSEs among different receiver trajectory control methods.}
\label{fig:PosRMSE_Opt}}
\end{figure}

\begin{figure}[htbp]
    \centering
{\includegraphics[width=0.35\textwidth]{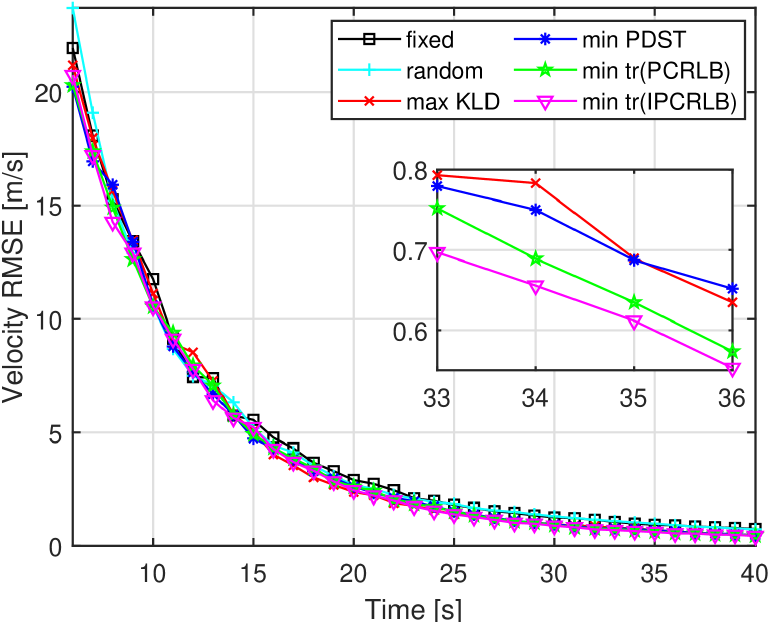}%
}
\caption{Comparison of velocity RMSEs among different receiver trajectory control methods.}
\label{fig:VelRMSE_Opt}
\end{figure}

The T2R geometries of different receiver trajectory control methods are shown in Fig.\ref{fig:OptReceiverTrajectory}. Except for the fixed and random methods, all others control the receiver moving towards the target but with different trajectories. Fig.\ref{fig:RkSNR_Opt} demonstrates the means and standard deviations of ${\sigma _d}$, ${\sigma _v}$, ${\sigma _\theta }$ and ${P_d}$ at each time among compared control methods, and the min tr(IPCRLB) delivers least TMU. After operating target tracking based upon acquired measurements, the RMSEs of position and velocity are given in Fig.\ref{fig:PosRMSE_Opt} and Fig.\ref{fig:VelRMSE_Opt}. Among compared control methods, since the min tr(IPCRLB) delivers least TMU, it provides the most accurate estimation of target position and velocity, followed by min tr(PCRLB), min (PDST) and max (KLD). The fixed and random methods perform the worst. Since IPCRLB fully accounts for the impact of T2R geometry on the TMU and eventual MSE lower bound, it formulates the bound more accurately than PCRLB for radar tracking in clutter, hence,  the min tr(IPCRLB) tends to optimize the T2R geometry more effectively than the min tr(PCRLB). Unlike the two bound-based control methods that optimize the T2R geometry to straightforwardly minimize theoretical tracking error, the min (PDST) minimizes the target-to-receiver distance and the max (KLD) maximizes information gain to control the receiver trajectory. Therefore, both min (PDST) and max (KLD) belong to indirect T2R geometry optimization, which inevitably results in degraded optimization benefits.

\section{Conclusion}\label{sec:6}
This paper investigates radar tracking in clutter with T2R geometry-dependent TMU. First of all, the target measurement error covariance is formulated as a function of transmitted signal, target and radar state, and its generalized form is then approximated to specify the impact of T2R geometry on error covariance. Based on the approximated uncertainty model, an improved PCRLB (IPCRLB) is rigorously derived for radar tracking in clutter by fully accounting for radar measurement uncertainty in terms of both MOU and geometry-dependent TMU. The derived IPCRLB is then applied to radar trajectory control to effectively optimize the T2R geometry, thereby acquiring high-quality target measurement for improved tracking accuracy. Numerical results validate that the new bound is more accurate than existing PCRLBs and provides an effective tool for sensor resource management. 

\section*{Appendix A} \label{appedix:A}
\section*{Derivation of  ${\delta _k}$ and $\cos \left({\frac{{{\beta _k}}}{2}} \right)$}

By a canonical definition \cite{ref:36}, the Doppler shift ${\xi_k}$ ignoring relativistic effects, is the time rate of change of bistatic range normalized by the wavelength $\frac{c}{{{f_c}}}$, i.e.,
\begin{equation}
{\xi _k} = \frac{{{f_c}}}{c}\left( {\frac{{d{R_{T,k}}}}{{dt}} + \frac{{d{R_{R,k}}}}{{dt}}} \right) \tag{A-1}
\end{equation}
When the target is moving (i.e., ${\left\| {{\bf{x}}_k^v} \right\|_2} \ne 0$) and the transmitter and receiver are stationary (${\left\| {{\bf{x}}_{T,k}^v} \right\|_2} = {\left\| {{\bf{x}}_{R,k}^v} \right\|_2} = 0$), the Doppler shift at the receiver due to the target motion is the sum of the projection of target velocity vector onto the transmitter-to-target line-of-sight (LOS) and the receiver-to-target LOS, normalized by the wavelength, given by
\begin{equation} \label{eq:A-2}
\begin{split}
& {\xi _{Tgt,k}} \\
&= \frac{{{f_c}}}{c}\left( {{{\left\| {{\bf{x}}_k^v} \right\|}_2}\cos \left( {{\delta _k} - \frac{{{\beta _k}}}{2}} \right) + {{\left\| {{\bf{x}}_k^v} \right\|}_2}\cos \left( {{\delta _k} + \frac{{{\beta _k}}}{2}} \right)} \right) \\
& = \frac{{2{f_c}}}{c}{\left\| {{\bf{x}}_k^v} \right\|_2}\cos \left( {{\delta _k}} \right)\cos \left( {\frac{{{\beta _k}}}{2}} \right) 
\end{split} \tag{A-2}
\end{equation}
As shown in Fig.\ref{fig:BisGeometry},   ${\delta _k}$ is the aspect angle of target velocity vector ${{\bf{x}}_k^v}$ referenced to the bistatic bisector,  measured positive clockwise from the bistatic bisector. When the target is stationary ( ${\left\| {{\bf{x}}_k^v} \right\|_2} = 0$) and the transmitter and receiver are moving (${\left\| {{\bf{x}}_{T,k}^v} \right\|_2} = {\left\| {{\bf{x}}_{R,k}^v} \right\|_2} \ne 0$), the Doppler shift at the receiver due to the combined transmitter and receiver motion is the sum of the projection of transmitter velocity vector onto transmitter-to-target LOS and receiver velocity vector onto receiver-to-target LOS, given by
\begin{equation} \label{eq:A-3}
\begin{split}
& {\xi _{TR,k}}  = \\
& \frac{{{f_c}}}{c}\left( {{{\left\| {{\bf{x}}_{T,k}^v} \right\|}_2}\cos \left( {{\delta _{T,k}} - {\theta _{T,k}}} \right) + {{\left\| {{\bf{x}}_{R,k}^v} \right\|}_2}\cos \left( {{\delta _{R,k}} - {\theta _k}} \right)} \right)
\end{split} \tag{A-3}
\end{equation}
where ${\delta _{T,k}}$ and ${\delta _{R,k}}$ are the aspect angles of transmitter velocity vector and receiver velocity vector referenced to the $x$-axis respectively, which are measured positive counter-clockwise from the $x$-axis. When the target, transmitter and receiver are all moving, the Doppler shift at the receiver due to all sites motion is the sum of (\ref{eq:A-2}) and (\ref{eq:A-3}), thus 
\begin{equation} \label{eq:A-4}
\begin{split}
{\xi _k} & =  \frac{{{f_c}}}{c}\left( {2{{\left\| {{\bf{x}}_k^v} \right\|}_2}\cos {\delta _k}\cos \left( {\frac{{{\beta _k}}}{2}} \right)} \right. \\ & \left.{+ {{\left\| {{\bf{x}}_{T,k}^v} \right\|}_2}\cos \left( {{\delta _{T,k}} - {\theta _{T,k}}} \right) + {{\left\| {{\bf{x}}_{R,k}^v} \right\|}_2}\cos \left( {{\delta _{R,k}} - {\theta _k}} \right)} \right)
\end{split} \tag{A-4}
\end{equation}
with the law of cosine, we have 
\begin{equation} \label{eq:A-5}
\begin{split}
\cos \left( {\frac{{{\beta _k}}}{2}} \right) &  = \sqrt {\frac{{1{\rm{ + }}\cos {\beta _k}}}{2}} \\ 
& = \sqrt {\frac{1}{2} + \frac{1}{2}\frac{{R_{T,k}^2 + R_{R,k}^2 - L_k^2}}{{2{R_{T,k}}{R_{R,k}}}}}
\end{split} \tag{A-5}
\end{equation}
As seen above, the Doppler shift ${\xi _k}$ implicitly relates to the bistatic range $d_k$ through the term $\cos \left( {\frac{{{\beta _k}}}{2}} \right)$, with $d_k = R_{T,k} + R_{R,k}$. Again, applying the law of cosines to the T2R triangle depicted in Fig.\ref{fig:BisGeometry}, with some mathematical transformations, we have
\begin{equation} \label{eq:A-6}
\begin{array}{l}
{R_{R,k}} = \frac{{d_k^2 - L_k^2}}{{2\left( {{d_k} + {L_k}\cos \left( {{\theta _k} - {\theta _{TR,k}}} \right)} \right)}}\\
{R_{T,k}} = \frac{{d_k^2 + L_k^2 + 2{d_k}{L_k}\cos \left( {{\theta _k} - {\theta _{TR,k}}} \right)}}{{2\left( {{d_k} + {L_k}\cos \left( {{\theta _k} - {\theta _{TR,k}}} \right)} \right)}}
\end{array} \tag{A-6}
\end{equation}
By substituting (\ref{eq:A-6}) to (\ref{eq:A-5}), $\cos \left({\frac{{{\beta _k}}}{2}} \right)$ can be explicitly expressed as a function of $d_k$ as shown in (5). Then by substituting (5) to (\ref{eq:A-4}), the mathematical relationship between $\xi_k$ and $d_k$ can be finally obtained as (4).
\section*{Appendix B}\label{appedix:B}
\section*{Derivation of ${\Lambda}_k^2({\bf{x}}_k,{m_k})$}
With the changed variable ${\bf{\tilde z}}_k^i$ defined in (40) and its corresponding integral region presented in (41), (36) can be rearranged as
\begin{equation}\label{eq:Lamda_k2Decomp}
\begin{split}
& {\rm{\Lambda }}_k^2({{\bf{x}}_k},{m_k}) \\
& = {{\rm{E}}_{{\bf{\tilde Z}}(k)}}\left\{ {{\Gamma _{{\bf{B}}{{\bf{B}}^T}}}} \right\}{\rm{ + }}{{\rm{E}}_{{\bf{\tilde Z}}(k)}}\left\{ {{\Gamma _{{\bf{B}}{{\bf{C}}^T}}}} \right\}{\rm{ + }}{{\rm{E}}_{{\bf{\tilde Z}}(k)}}\left\{ {{\Gamma _{{\bf{C}}{{\bf{C}}^T}}}} \right\} 
\end{split}\tag{B-1}
\end{equation}
with each expectation terms above given by (\ref{eq:E_BB_Orig}), (\ref{eq:E_BC_Orig}), (\ref{eq:E_CC_Orig}), let define
\begin{figure*}[htb]
\
\hrulefill
\vspace*{2pt}
\centering
\begin{equation}\label{eq:E_BB_Orig}
\begin{split}
{{\rm{E}}_{{\bf{\tilde Z}}(k)}}\left\{ {{\Gamma _{{\bf{B}}{{\bf{B}}^T}}}} \right\} & = \frac{{{\lambda ^2}V_g^2m_k^2{{\left( {P_{FA}^g} \right)}^{2/\left( {1 + {\Psi _k}({{\bf{x}}_k})} \right)}}}}{{{\ell_g^4}}}\frac{{{{\ln }^2}P_{FA}^g}}{{{{[1 + {\Psi _k}({{\bf{x}}_k})]}^4}}} \\
& \times \int\limits_{{\bf{\tilde z}}_k^1 \in \hat \Xi } { \cdots \int\limits_{{\bf{\tilde z}}_k^{{m_k}} \in \hat \Xi } {\frac{1}{{p({\bf{\tilde Z}}(k)|{{\bf{x}}_k},{m_k})}}{\underbrace {\left[ {\frac{{\sum\limits_{i = 1}^{{m_k}} {\sum\limits_{j = 1}^{{m_k}} {p({\bf{\tilde z}}_k^i)p({\bf{\tilde z}}_k^j)} } }}{{m_k^2V_g^{2{m_k} - 2}}} - \frac{{2\sum\limits_{i = 1}^{{m_k}} {p({\bf{\tilde z}}_k^i)} }}{{{m_k}V_g^{2{m_k} - 1}}} + \frac{1}{{V_g^{2{m_k}}}}} \right]}_{{\Omega _1}}}{\rm{d}}{\bf{\tilde z}}_k^1 \cdots {\rm{d}}{\bf{\tilde z}}_k^{{m_k}}} } 
\end{split}\tag{B-2}
\end{equation}
\hrulefill
\vspace*{-2pt}
\end{figure*}
\begin{figure*}[htb]
\centering
\begin{equation}\label{eq:E_BC_Orig}
\begin{split}
{{\rm{E}}_{{\bf{\tilde Z}}(k)}}\left\{ {{\Gamma _{{\bf{B}}{{\bf{C}}^T}}}} \right\} & = \frac{{ {d_g}({{\bf{x}}_k},{m_k})}}{{\Psi_k ({{\bf{x}}_k})V_g^{{m_k} - 2}}}\frac{{\ln P_{FA}^g}}{{{{[1 + {\Psi _k}({{\bf{x}}_k})]}^2}}}\frac{{\lambda {{\left( {P_{FA}^g} \right)}^{1/\left( {1 + {\Psi _k}({{\bf{x}}_k})} \right)}}}}{{\ell _g^2}} \\
& \times \int\limits_{{\bf{\tilde z}}_k^1 \in \hat \Xi } { \cdots \int\limits_{{\bf{\tilde z}}_k^{{m_k}} \in \hat \Xi } {\frac{1}{{p({\bf{\tilde Z}}(k)|{{\bf{x}}_k},{m_k})}}{\underbrace {\left[ {\frac{{\sum\limits_{i = 1}^{{m_k}} {\sum\limits_{j = 1}^{{m_k}} {p({\bf{\tilde z}}_k^i)} p({\bf{\tilde z}}_k^j)\left[ {\tilde \gamma \left( {{\bf{\tilde z}}_k^j} \right) - n} \right]} }}{{{m_k}V_g^{{m_k} - 1}}} - \frac{{\sum\limits_{i = 1}^{{m_k}} {p({\bf{\tilde z}}_k^i)\left[ {\tilde \gamma \left( {{\bf{\tilde z}}_k^i} \right) - n} \right]} }}{{V_g^{{m_k}}}}} \right]}_{{\Omega _2}}}{\bf{\tilde z}}_k^1 \cdots {\rm{d}}{\bf{\tilde z}}_k^{{m_k}}} }
\end{split}\tag{B-3}
\end{equation}
\hrulefill
\vspace*{-2pt}
\end{figure*}
\begin{figure*}[htb]
\small
\centering
\begin{equation}\label{eq:E_CC_Orig}
{{\rm{E}}_{{\bf{\tilde Z}}(k)}}\left\{ {{\Gamma _{{\bf{C}}{{\bf{C}}^T}}}} \right\} = \frac{{d_g^2({{\bf{x}}_k},{m_k})}}{{4{\Psi_k^2}({{\bf{x}}_k})m_k^2V_g^{2{m_k} - 2}}}\int\limits_{{\bf{\tilde z}}_k^1 \in \hat \Xi } { \cdots \int\limits_{{\bf{\tilde z}}_k^{{m_k}} \in \hat \Xi } {\frac{1}{{p({\bf{\tilde Z}}(k)|{{\bf{x}}_k},{m_k})}}{\underbrace {\sum\limits_{i = 1}^{{m_k}} {\sum\limits_{j = 1}^{{m_k}} {p({\bf{\tilde z}}_k^i)} \left[ {\tilde \gamma \left( {{\bf{\tilde z}}_k^i} \right) - n} \right]p({\bf{\tilde z}}_k^j)\left[ {\tilde \gamma \left( {{\bf{\tilde z}}_k^j} \right) - n} \right]} }_{{\Omega _3}}}{\rm{d}}{\bf{\tilde z}}_k^1 \cdots {\rm{d}}{\bf{\tilde z}}_k^{{m_k}}} } \tag{B-4}
\end{equation}
\hrulefill
\vspace*{-2pt}
\end{figure*}
\begin{equation}
{\ell _g} = \left[ {1 - P_d^g({{\bf{x}}_k})} \right]\lambda {V_g} + {m_k}P_d^g({{\bf{x}}_k}) \tag{B-5}
\end{equation}
and
\begin{equation}\label{eq:B-6}
\tilde \gamma \left( {{\bf{\tilde z}}_k^i} \right) = {\left( {{\bf{\tilde z}}_k^i} \right)^T}{\bf{R}}_k^{ - 1}({{\bf{x}}_k}){\bf{\tilde z}}_k^i \tag{B-6}
\end{equation}
It is easy to see from (\ref{eq:E_BB_Orig}) that the integral of ${p({\bf{\tilde z}}_k^i)p({\bf{\tilde z}}_k^j)}$ is all equal in value for $i=1,...,m_k$, $j=1,...,m_k$, and the integral of ${p({\bf{\tilde z}}_k^i)}$ is also all equal in value for $i=1,...,m_k$, therefore, by removing the summation notations inside the integral function of (\ref{eq:E_BB_Orig}), ${{\Omega _1}}$ reduces to
\begin{equation}\label{eq:Omega_1}
\begin{split}
{\Omega _1} = &  \frac{{{e^{ - \tilde \gamma \left( {{\bf{\tilde z}}_k^1} \right)}}{\rm{ + }}\left( {{m_k} - 1} \right){e^{ - \frac{{\tilde \gamma \left( {{\bf{\tilde z}}_k^1} \right) + \tilde \gamma \left( {{\bf{\tilde z}}_k^2} \right)}}{2}}}}}{{{m_k}V_g^{2{m_k} - 2}{{\left( {2\pi } \right)}^n}\left| {{{\bf{R}}_k}\left( {{{\bf{x}}_k}} \right)} \right|}}  \\
&- \frac{{2{e^{ - \frac{{\tilde \gamma \left( {{\bf{\tilde z}}_k^1} \right)}}{2}}}}}{{V_g^{2{m_k} - 1}\sqrt {{{\left( {2\pi } \right)}^n}\left| {{{\bf{R}}_k}\left( {{{\bf{x}}_k}} \right)} \right|} }} + \frac{1}{{V_g^{2{m_k}}}}
\end{split}\tag{B-7}
\end{equation}
Similarly, by removing the summation notations inside the integral function of (\ref{eq:E_BC_Orig}) and (\ref{eq:E_CC_Orig}), we have
\begin{equation}\label{eq:Omega_2}
\begin{split}
&{\Omega _2} = - \frac{{{m_k}{e^{ - \frac{{\tilde \gamma \left( {{\bf{\tilde z}}_k^1} \right)}}{2}}}\left[ {\tilde \gamma \left( {{\bf{\tilde z}}_k^1} \right) - n} \right]}}{{V_g^{{m_k}}\sqrt {{{\left( {2\pi } \right)}^n}\left| {{{\bf{R}}_k}\left( {{{\bf{x}}_k}} \right)} \right|} }} + \\
& \frac{{{e^{ - \tilde \gamma \left( {{\bf{\tilde z}}_k^1} \right)}}\left[ {\tilde \gamma \left( {{\bf{\tilde z}}_k^1} \right) - n} \right] + \left( {{m_k} - 1} \right){e^{ - \frac{{\tilde \gamma \left( {{\bf{\tilde z}}_k^1} \right) + \tilde \gamma \left( {{\bf{\tilde z}}_k^2} \right)}}{2}}}\left[ {\tilde \gamma \left( {{\bf{\tilde z}}_k^2} \right) - n} \right]}}{{V_g^{{m_k} - 1}{{\left( {2\pi } \right)}^n}\left| {{{\bf{R}}_k}\left( {{{\bf{x}}_k}} \right)} \right|}}
\end{split}\tag{B-8}
\end{equation}
\begin{equation}\label{eq:Omega_3}
\begin{split}
& {\Omega _3} = \frac{{{m_k}{e^{ - \tilde \gamma \left( {{\bf{\tilde z}}_k^1} \right)}}{{\left[ {\tilde \gamma \left( {{\bf{\tilde z}}_k^1} \right) - n} \right]}^2}}}{{{{\left( {2\pi } \right)}^n}\left| {{{\bf{R}}_k}\left( {{{\bf{x}}_k}} \right)} \right|}} +  \\
& \frac{{{m_k}\left( {{m_k} - 1} \right){e^{ - \frac{{\tilde \gamma \left( {{\bf{\tilde z}}_k^1} \right) + \tilde \gamma \left( {{\bf{\tilde z}}_k^2} \right)}}{2}}}\left[ {\tilde \gamma \left( {{\bf{\tilde z}}_k^1} \right) - n} \right]\left[ {\tilde \gamma \left( {{\bf{\tilde z}}_k^2} \right) - n} \right]}}{{{{\left( {2\pi } \right)}^n}\left| {{{\bf{R}}_k}\left( {{{\bf{x}}_k}} \right)} \right|}}
\end{split}\tag{B-9}
\end{equation}
To further simplify the integral functions above, again make the change of variable
\begin{equation}\label{eq:zki_hat}
{\bf{\hat z}}_k^i[h] = \frac{{{\bf{\tilde z}}_k^i[h]}}{{{\sigma _h}}}, h = 1, \cdots ,n \tag{B-10}
\end{equation}
let ${\bf{\hat z}}_k^i = {\left[ {{\bf{\hat z}}_k^i[1]{\kern 1pt} {\kern 1pt} {\kern 1pt} {\kern 1pt} {\kern 1pt} {\kern 1pt}  \cdots {\kern 1pt} {\kern 1pt} {\kern 1pt} {\kern 1pt} {\kern 1pt} {\bf{\hat z}}_k^i[n]} \right]^T}{\rm{,}}{\kern 1pt} {\kern 1pt} {\kern 1pt} {\kern 1pt} i = 1, \ldots ,{m_k}$,  the integral region of ${\bf{\hat z}}_k^i$ then converts to $\mathord{\buildrel{\lower3pt\hbox{$\scriptscriptstyle\smile$}} 
\over \Xi }  = \left[ { - g,{\kern 1pt} g} \right] \times  \cdots  \times \left[ { - g,{\kern 1pt} g} \right]$, hence we have the following differentiating relationship
\begin{equation}\label{eq:diff_zk1_Tilde}
{\rm{d}}{\bf{\tilde z}}_k^1 \cdots {\rm{d}}{\bf{\tilde z}}_k^{{m_k}} = {\left| {{{\bf{R}}_k}({{\bf{x}}_k})} \right|^{{m_k}/2}}{\rm{d}}{\bf{\hat z}}_k^1 \cdots {\rm{d}}{\bf{\hat z}}_k^{{m_k}} \tag{B-11}
\end{equation}
and $\tilde \gamma \left( {{\bf{\tilde z}}_k^i} \right)$ in (\ref{eq:B-6}) can be simplified as
\begin{equation}\label{eq:gamma_tilde}
\tilde \gamma \left( {{\bf{\tilde z}}_k^i} \right) = \sum\limits_{h = 1}^n {{{\left( {{\bf{\hat z}}_k^i[h]} \right)}^2}}  \tag{B-11}
\end{equation}
Substituting (\ref{eq:gamma_tilde}) to (\ref{eq:Omega_1}) , (\ref{eq:Omega_2}), (\ref{eq:Omega_3}), their results together with (\ref{eq:diff_zk1_Tilde}) are then substituted to (\ref{eq:E_BB_Orig})-(\ref{eq:E_CC_Orig}), the final form of  (\ref{eq:Lamda_k2Decomp}) can be obtained as in (43)-(46).

\bibliography{mybibfile}

\end{document}